# Hourly evolution of intra-urban temperature variability across the local climate zones. The case of Madrid

Miguel Núñez-Peiró[a,*]; C. Sánchez-Guevara Sánchez[a]; F. Javier Neila González[a]

[a] Escuela Técnica Superior de Arquitectura, Universidad Politécnica de Madrid. Avda. Juan de Herrera 4, 28040, Madrid, Spain.

[*] Corresponding author. E-mail address: miguel.nunez@upm.es (M. Núñez-Peiró)

**Abstract**

Field measurement campaigns have grown exponentially in recent years, stemming from the need for reliable data to validate urban climate models and obtain a better understanding of urban climate features. Also contributing to this growth is the Local Climate Zone (LCZ) scheme, firstly developed to enhance the accuracy in the contextualisation of urban measurements, and lately used for characterising urban areas. Due to its relative novelty, researchers are still investigating the potential of LCZs and its indicators for urban temperature variability detection. In this respect, the present study introduces the results of an extensive monitoring campaign carried out in the city of Madrid over a two-year period (2016-2018). The aim of this work is to further examine the relationships between LCZs and air temperature differences, with emphasis on their hourly and seasonal evolution. A graphical and statistical analysis to identify temperature variability trends for each LCZ is performed. Results support the existing evidence suggesting a high level of effectiveness in capturing the heat island (UHI) profile of different urban areas, while underperforming when it comes to capturing diurnal temperature variability. The incorporation of indicators that explain the daytime temperature variation phenomenon into the LCZ scheme is therefore recommended, warranting further research.

1. **Introduction**

Climate and cities are deeply interconnected. For centuries, cities have adapted to the climatic conditions of their environment. The way in which climate influences cities can be easily noticed through adaptive morphological actions such as street narrowing or roof pitch increases. Today, climate change is urging cities to adapt and adopt new strategies for facing global warming from an integrated urban perspective (Reckien et al., 2018). Similarly, the effect of cities on climate conditions is also well known: not only do they contribute to global warming (Dodman, 2009), but they also modify meso- and micro-climatic conditions (Oke, 1982; Oke et al., 2017a).

From a historical perspective (e.g. Hebbert, 2014; Mills, 2014; Stewart, 2019), the roots of urban climatology go back to the 19$^{th}$ century, when the first studies on air pollution and urban temperature differences were published (Howard, 1833; Renou, 1858; Rusel, 1888). Since then, urban climate studies evolved from urban-rural comparisons of multiple meteorological variables (e.g. Chandler, 1965; Geiger, 1950; Kratzer, 1937) to complex urban land surface models that can be coupled with meso-climatic (Ching, 2013; Jandaghian and Berardi, 2020) and building energy models (Lauzet et al., 2019; Mirzaei, 2015). Nowadays, the latter are also being used for in-depth exploration and unravelling of the most complex climatic processes at the urban scale, which would otherwise be impossible to discern in a real urban environment. Some recent examples are the study of the anthropogenic heat dispersion (Doan et al., 2019; Yuan et al., 2020), the effect of water bodies on the environment (Ampatzidis and Kershaw, 2020), or the relationship between heat loads and social inequalities (Zuvela-Aloise, 2017).

Despite the shift in focus from observation to modelling, measurements remain pivotal in urban climate research. Urban climate models still present a high level of uncertainty, as well as significant heterogeneity between them (Grimmond et al., 2011, 2010). They have yet to be tested under different urban contexts, in particular at the micro-climatic scale, to prove their reliability (Best and Grimmmmond, 2015; Toparlar et al., 2017). Measurement campaigns are, therefore, required for validating their performance (Velasco, 2018). On-site observations might also provide an improved understanding of the dynamic processes that govern the urban climate, leading to novel theoretical and modelling approaches, or the improvement of the existing ones (Barlow, 2014; Karl et al., 2020). On top of this, experimental urban climatic data can be used for a variety of multidisciplinary purposes, such as evaluating population vulnerability towards high temperatures (Jänicke et al., 2018; López-Bueno et al., 2019, 2020; Sánchez-Guevara et al., 2019; Willers et al., 2016), assessing buildings' energy consumption (Kolokotroni et al., 2012; Pyrgou et al., 2017), or investigating urban phenological patterns (i.e. pollen production, Jianan et al., 2007; Jochner et al., 2011).

*1.1.    Using Local Climate Zones for contextualising and characterising urban areas*

The Local Climate Zone (LCZ) scheme is a climate classification system for urban environments proposed by Stewart and Oke (2012). LCZs aim at clustering urban (and rural) contexts into 17 conceptual units (10 built-up and 7 land cover types), each one representing their unique local thermal characteristics. A set of 10 quantitative parameters linked to their morphology (sky view factor, aspect ratio, height of roughness elements, terrain roughness class), the surface cover (building, impervious and pervious surface fraction) and their thermal, radiative and metabolic properties (surface admittance, surface albedo, anthropogenic heat output) differentiate each unit.

Although the original purpose of the LCZs was to strengthen the contextualisation and inter-comparability of urban temperature measurements, their use has extended to contextualising other measured parameters, such as $PM_{2.5}$ (Shi et al., 2019), VOCs (Valach et al., 2015), $CH_4$ (Pawlak and Fortuniak, 2016), $CO_2$ (Christen, 2014; Crawford and Christen, 2015; Kurppa et al., 2015; Menzer and McFadden, 2017; Roth et al., 2016; Velasco et al., 2014), and energy fluxes (Ando and Ueyama, 2017; Feigenwinter et al., 2012; Kotthaus and Grimmond, 2014). LCZs have demonstrated to be useful for other purposes as well, namely providing climatic guidance for urban planning (Alexander et al., 2016; Perera and Emmanuel, 2016) and modelling weather conditions in urban environments (Brousse et al., 2016; Hammerberg et al., 2018). In just a few years, the LCZs have been rapidly and intensively adopted, becoming the standard scheme in urban climate description.

Regarding the scheme's performance in urban micro-climatic characteristics detection, several studies have shown that its different classes tend to exhibit distinctive temperature profiles (see **Table 1**). In previous studies, the preferred methodology for obtaining temperature data is in-situ monitoring campaigns, both fixed and mobile. Data from urban models and Citizen Weather Stations (CWS) are also popular, although concerns over their reliability remain (Bell et al., 2015; Chapman et al., 2017; Gardes et al., 2020; Kwok et al., 2019). Some of these studies have also pointed out the need for assessing the performance of the LCZs, rather than just confirming that they display different trends. More specifically, they have focused on evaluating whether the differences between LCZs are statistically significant (Beck et al., 2018; Fenner et al., 2017; Leconte et al., 2020; Richard et al., 2018), if they concentrate at certain times of the year or under specific meteorological conditions (Arnds et al., 2017; Thomas et al., 2014; Yang et al., 2018), or if other parameters might affect the LCZs inter- and intra-variability (Kotharkar et al., 2019; Kwok et al., 2019; Leconte et al., 2017). However, research on the topic is still scarce and limited to a specific climatic context (mostly *Cf*, humid and warm temperate climates), and in some cases is based on short datasets. It thus seems necessary to continue investigating these links and expand the scope to include other climatic contexts.

## 1.2. *Aim of the study*

This study presents the results of an extensive monitoring campaign carried out in the city of Madrid over a two-year period (2016-2018). This work aims to further examine the relationships between LCZs and ambient temperature differences, with emphasis on their hourly and seasonal evolution. For that purpose, a comprehensive overview of the collected data and its associated metadata is firstly provided, describing the pre-processing techniques (i.e. Quality Control (QC) procedures) and the LCZs' contextualisation of the urban measurements. Then, a graphical analysis to identify the trends of the air temperature, Urban Heat Island (UHI) intensity, and cooling rate profiles, is performed for each LCZ. Finally, differences between the LCZs, as well as the ability of the LCZ indicators to capture urban temperature differences, are statistically evaluated on an hourly basis and at different times of the year.

**Table 1.** Previous cities in which urban temperature variability across different LCZs was compared. It includes the background climate, the source used to obtain air temperature data, and the length of the dataset.

| City | K-G Climate [1] | Reference | On-site air temperature measurements | | | | Air temp. modelling | Length of the dataset [4] |
|---|---|---|---|---|---|---|---|---|
| | | | Fixed official [2] | Fixed non-official [2] | CWS [3] | Transects | | |
| Delhi (India) | Cwa – BSh | Budhiraja et al. (2020) | ■ | | | | | 5 days (CH, May 2018) |
| Nancy (France) | Cfb | Leconte et al. (2020, 2015, 2017) | | | | ■ | | 2 days (D, Aug 2013) |
| Nanjing (China) | Cfa | Yang et al. (2020b, 2018) | | ■ | | | | 3 years (CH, Aug 2016 – Jul 2019) |
| Sendai (Japan) | Cfa | Zhou et al. (2020) | | ■ | | | | 11 days (CH, Aug 2018) |
| - (France) [5] | Csa – Cfb | Gardes et al. (2020) | | | | | ■ | 6 days (CH) |
| Nagpur (India) | Aw – As | Kotharkar et al. (2019) | | ■ | | ■ | | 6 days (CH, D, Apr 2016, Mar – Apr 2017) |
| | | Kotharkar and Bagade (2018) | | ■ | | ■ | | 5 days (CH, D, Dec 2015 – Feb 2016) |
| Novi Sad (Serbia) | Cfa | Šećerov et al. (2019) | | ■ | | | | 2 months (CH, Jun – Aug 2015) |
| Toulouse (France) | Cfa | Kwok et al. (2019) | | ■ | | | ■ | 18 days (CH, Jun – Aug 2004) |
| Vienna (Austria) | Cfb | Hammerberg et al. (2018) | ■ | | ■ | | ■ | 10 days (CH, Jan – Jul 2015) |
| Antwerp (Belgium) | Cfb | Verdonck et al. (2018) | | ■ | | | ■ | 82 days (D, Jun – Aug 2014, 2015) |
| Brussels (Belgium) | Cfb | Verdonck et al. (2018) | | ■ | | | ■ | 101 days (D, Jun - Aug 2014, 2015) |
| Ghent (Belgium) | Cfb | Verdonck et al. (2018) | | ■ | | | ■ | 76 days (D, Jun – Aug 2014, 2015) |
| Augsburg (Germany) | Cfb | Beck et al. (2018) | | ■ | | | | ~3 years (CH, Dec 2014 – Oct 2017) |
| | | Verdonck et al. (2018) | | ■ | | | ■ | ~8 months (CH, Jun – Sep 2014, 2015) |
| Dijon (France) | Cfb | Richard et al. (2018) | | ■ | | | ■ | 3 weeks (CH, Jul 2015) |
| Berlin (Germany) | Cfb | Fenner et al. (2017) | | | ■ | | | 12 months (CH, Jan – Dec 2015) |
| | | Fenner et al. (2014) | | ■ | | | | 10 years (CH, 2001 – 2010) |
| Hamburg (Germany) | Cfb | Arnds et al. (2017) | ■ | ■ | | | | 27 years (CD, 1985 – 2012) |
| Matsuyama (Japan) | Cfa | Thapa Chhetri et al. (2017) | | ■ | | | | 6 days (CH, Aug) |
| Szeged (Hungary) | Cfa | Skarbit et al. (2017) | | ■ | | | | 1 year (CH, Jun 2014 – May 2015) |
| | | Unger et al. (2015) | | ■ | | | | 2 days (CH, Mar 2014) |
| | | Lelovics et al. (2014) | | | | ■ | | 4 days (D, Apr 2002 – Mar 2003) |
| Olomouc (Czech Rep) | Cfb | Lehnert et al. (2015) | | ■ | | | | 15 days (CH, Jul 2010 – Oct 2011) |
| Dublin (Ireland) | Cfb | Alexander & Mills (2014) | | ■ | | ■ | | 7 days (CH, D, Aug – Sep 2010) |
| Kochi (India) | Am | Thomas et al. (2014) | | | | ■ | | 7 days (D, Jan 2011 – Mar 2013) |
| Nagano (Japan) | Cfa – Dfb | Stewart et al. (2014) | | | | ■ | | 32 days (D, Dec 2001 – Nov 2002) |
| Uppsala (Sweden) | Cfb | Stewart et al. (2014) | | ■ | | ■ | | 31 days (D, ~1950); 3 days (CH, Sep 1976) |
| Vancouver (Canada) | Cfb – Csb | Stewart et al. (2014) | | | | ■ | | 5 days (D, Nov 1999, March 2008, 2010) |
| Hong Kong (China) | Cwa | Siu & Hart (2013) | ■ | | | | | 20 years (CD, 1989 – 2008; CH, 2004 – 2008) |
| Mendoza (Argentina) | BWk – BWh | Puliafito et al. (2013) | | | | ■ | | ~5 days (D, Dec 2004 – Feb 2005) |
| Glasgow (UK) | Cfb | Emmanuel & Krüger (2012) | ■ | | | | | 50 years (CH, 1959 – 2009) |

[1] According to the updated version of the Köppen-Geiger climatic maps shown in (Kottek et al., 2006; Rubel et al., 2017). [2] "official" refers to those measurements derived from official meteorological or air quality networks. [3] CWS: Citizen Weather Stations. [4] CH: Continuous Hourly observations; CD: Continuous Daily observations; D: Discrete observations. [5] Includes several urban areas.

## 2. Materials and methods

### 2.1. Study area

The present study focuses on the city of Madrid, located in the centre of Spain (40.42 N, 3.70 W). According to the Köppen-Geiger classification (Kottek et al., 2006; Rubel et al., 2017), Madrid has a Mediterranean climate bordering the semi-arid class (*Csa – BSk*), with hot summers and cold winters. Precipitation tends to concentrate in spring and autumn, while cloudless days are mostly observed in summer. Regionally, its climate is slightly influenced by the presence of the Central System, a mountain range that divides the inner plateau of the Iberian Peninsula into two parts and influences the wind direction (mostly NE-SW), creating a north-south temperature gradient from colder to warmer conditions. The *Manzanares* river, which crosses the city from north to south, further contributes to this effect, channelling cold air from the mountains into the city (Fernández García et al., 1996).

With a population of 3.3 M inhabitants, 6.9 M if we consider the functional urban area (Eurostat, 2020), Madrid is the largest city of the country. It experienced an intense urban development during the 1960s and 1970s, when nearly 40% of the existing housing stock was built. Nowadays, Madrid presents a concentric radial distribution with a predominant north-south axis. The eastern border of the city is delimited by the metropolitan park *Casa de Campo* and the Mediterranean forest *Monte de El Pardo*, while the western side of the city gathers the bulk of the new urban developments. Overall, there is a prevalence of compact midrise (LCZ 2) and open midrise (LCZ 5) climatic zones. The large low-rise urban class (LCZ 8) is also substantially present in the south-western periphery (see **Fig. 1**).

### 2.2. Equipment and location

In cooperation with the municipality, 20 sensor units were deployed across the city of Madrid during the MODIFICA Project (Universidad Politécnica de Madrid, 2014, see **Fig. 1**). The distribution of the equipment followed a gradient approach (Muller et al., 2013), with a denser concentration of sensors in the city centre, taking into account the temperature gradient found in a previous study based on urban transects (Núñez Peiró et al., 2017). The campaign complied with the World Meteorological Organisation (WMO) guidelines for carrying out measurements within urban environments, in terms of sensor siting and metadata documentation (WMO, 2017a). This information is included in the **Appendix**, and an example can be seen in **Fig. 2**. Regarding the sensors siting, these were placed on lampposts 5-6 m above the ground. East-west streets were preferred for the sensors' location, to avoid daytime temperature peaks during the summer. In the absence of a standardised method to calculate source areas within the Urban Canopy Layer (UCL), an estimation based on a circumference of 500 m radius was used to analyse the urban structure homogeneity and guarantee the measurement representativeness (Núñez Peiró et al., 2019). Two locations, *Imperial* (site 2) and *Los Cármenes* (site 13) failed to meet this last requirement and were therefore excluded from this paper. This data was complemented with the official temperature data from 5 meteorological observatories belonging to the State Meteorological Agency (AEMET).

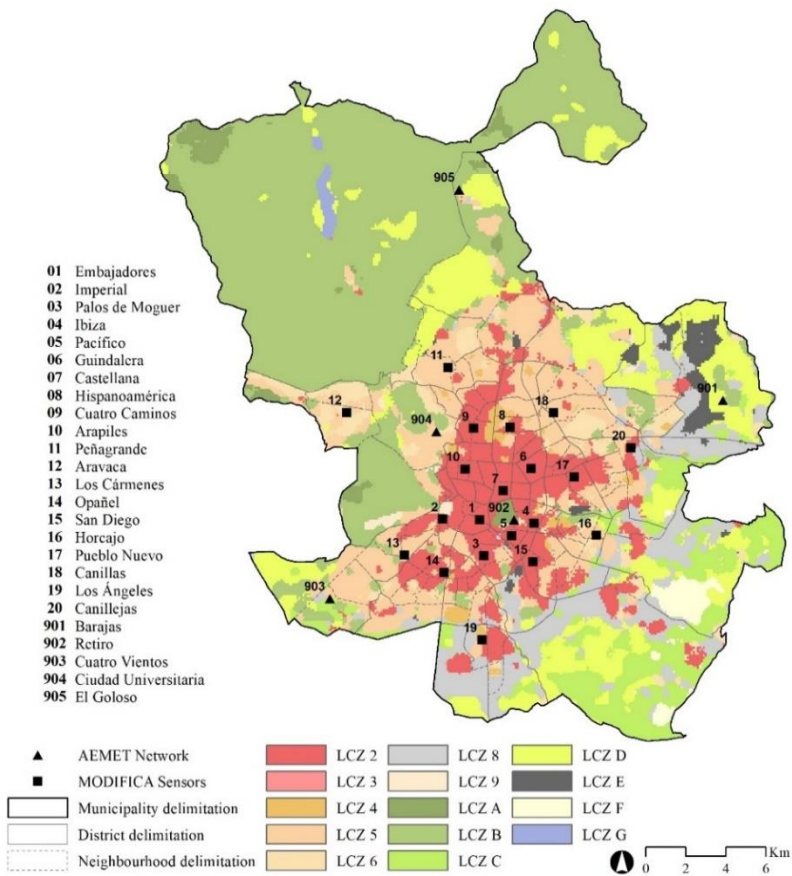

**Fig. 1.** Location of the MODIFICA sensors and the AEMET meteorological stations within the city of Madrid. The World Urban Database and Access Portal Tools (WUDAPT) LCZs (Brousse et al., 2016) are presented as a background layer.

The measurements were made with HOBO U23-001 temperature and relative humidity dataloggers. These have an accuracy of ±0.2 °C and ±2.5% for the temperature and the relative humidity, respectively. Many researchers have previously employed this equipment in UHI studies, using predominantly naturally ventilated radiation shields (e.g. Beck et al., 2018; Borbora and Das, 2014; Coseo and Larsen, 2014; Kotharkar and Bagade, 2018; Kourtidis et al., 2015; Richard et al., 2018; Schatz and Kucharik, 2014; Suomi, 2018; Yang et al., 2020b). In the present study, a bespoke mechanically ventilated, low-cost radiation shield was developed to improve the accuracy of the temperature measurements within the UCL during the daytime (Núñez Peiró et al., 2018; see the **Appendix**).

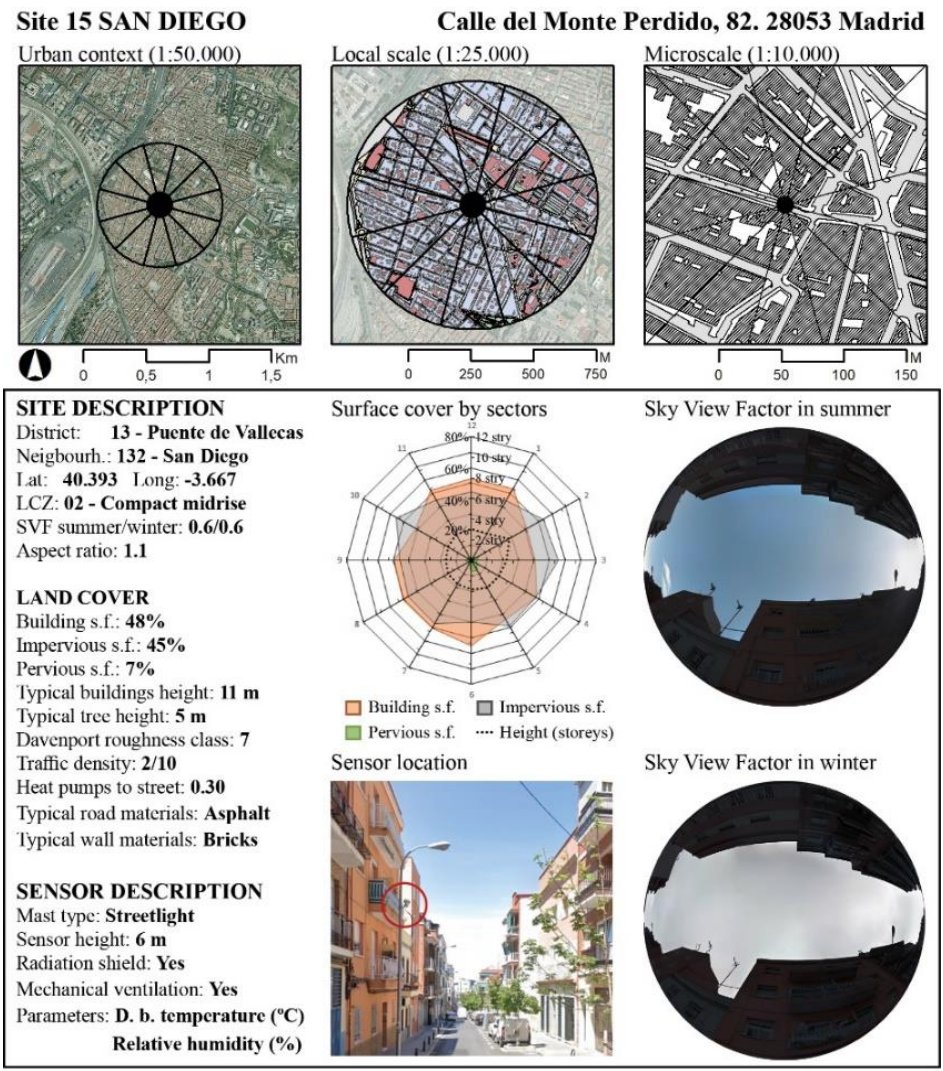

**Fig. 2.** Example of the metadata associated to each MODIFICA sensor, describing both the local scale and the micro-scale of the sensor located at the site #15 (San Diego), and including information of the measurement site and its surroundings. The metadata associated to the other sensors can be found in the Appendix.

The classification of each sensor within the LCZs was estimated manually to contrast the WUDAPT information (Bechtel et al., 2019, 2015). The surface cover and geometric parameters for the source area of the sensors were determined using the municipal cartography (Ayuntamiento de Madrid, 2015) and the national land registry (Ministerio de Hacienda, 2019). The aspect ratio (AR) was established for the street in which the sensor was located. The sky view factor (SVF) was calculated at the location of each sensor using Google Street View panorama images (Li et al., 2017; Miao et al., 2020). Since the changes in foliage were meaningful in most streets, SVFs were computed for both summer (SVFS) and winter (SVFW). The values for thermal, radiative, and metabolic properties could not be drawn due to the lack of available and reliable data sources. Instead, values for traffic density (Ayuntamiento de Madrid, 2013), the ratio of heat pumps per dwelling overlooking the street, typical road and wall materials data were included in the metadata. The MODIFICA network sites were classified as LCZ 2 (10), LCZ 4 (2), LCZ 5 (3), LCZ 6 (2), and LCZ 9 (1). The AEMET sites were classified as LCZ 9 (1), LCZ A (1), LCZ C (1), and LCZ D (2). **Table 2** shows a comparison between the estimated LCZs classification according to the manual classification procedure and the WUDAPT database. Overall, the manual classification revealed a higher LCZ variability than the WUDAPT, particularly on the city's periphery. Even though the discrepancies were not negligible, they mostly occurred with adjacent classes (e.g. LCZ 4 and 5).

**Table 2.** Classification of the measurement sites using the LCZ scheme. A manual classification based on the data summarised in the Appendix is compared with the WUDAPT classification.

| Classif. method | Site ID | | | | | | | | | | | | | | | | | | | | | | |
|---|---|---|---|---|---|---|---|---|---|---|---|---|---|---|---|---|---|---|---|---|---|---|---|
| | 01 | 03 | 04 | 05 | 06 | 07 | 08 | 09 | 10 | 11 | 12 | 14 | 15 | 16 | 17 | 18 | 19 | 20 | 901 | 902 | 903 | 904 | 905 |
| Manual | 2 | 2 | 4 | 2 | 2 | 2 | 2 | 2 | 2 | 4 | 6 | 5 | 2 | 9 | 2 | 6 | 5 | 5 | D | A | D | 9 | C |
| WUDAPT | 2 | 2 | 2 | 2 | 2 | 2 | 2 | 2 | 2 | 5 | 6 | 2 | 2 | 5 | 2 | 5 | 4 | 2 | D | A | C | 5 | C |

### *2.3.  Data and Quality Control procedures*

Two years of screen-height air temperature were collected between August 2016 and July 2018 on an hourly basis. The temperature series of the MODIFICA campaign were, for all sites, mostly complete. Only one significant discontinuity of 68 hours, due to technical issues, was registered in October 2017. Quality Control (QC) procedures were also applied to this dataset. These were derived from the WMO Guidelines for Level II data (WMO, 2017b, 2017a; Zahumensky, 2004), and included plausible value and time consistency checks. Additionally, a test for evaluating spatial consistency was conducted, analysing whether the gap between a specific record and its surrounding data was too large when compared to the average. If it remained within 4 standard deviations, it was not considered suspect data. Records were marked as erroneous when flagged as suspect twice. Regarding the temperature records provided by the AEMET, these only presented small discontinuities (<2 consecutive hours) and, since these records are subject to regular QC analysis before publishing, they were not included in the QC analysis. Neither erroneous nor missing records were replaced but were removed from the series. **Table 3** summarises the aforementioned process.

**Table 3.** Summary of the records obtained from the MODIFICA campaign and the AEMET observatories. Records were flagged as missing, suspect, erroneous or correct according to QC procedures.

| Sensor ID | Site name | Total data | Missing values | Flagged as suspect ||| Flagged as erroneous | Flagged as correct |
|---|---|---|---|---|---|---|---|---|
| | | | | Plausible value | Time consist. | Space consist. | | |
| **MODIFICA campaign** | | | | | | | | |
| 01 | Embajadores | 17452 | 68 | 0 | 40 | 24 | 3 | 17449 |
| 02 | Imperial | 17452 | 68 | 0 | 851 | 69 | 10 | 17442 |
| 03 | La Chopera | 17452 | 68 | 0 | 252 | 30 | 5 | 17447 |
| 04 | Estrella | 17452 | 68 | 0 | 56 | 22 | 5 | 17447 |
| 05 | Pacífico | 17453 | 67 | 0 | 29 | 13 | 2 | 17450 |
| 06 | Guindalera | 17452 | 68 | 0 | 40 | 24 | 3 | 17449 |
| 07 | Recoletos | 17452 | 68 | 0 | 36 | 18 | 2 | 17450 |
| 08 | Hispanoamérica | 17452 | 68 | 0 | 45 | 22 | 3 | 17449 |
| 09 | Cuatro Caminos | 17452 | 68 | 0 | 36 | 22 | 2 | 17450 |
| 10 | Arapiles | 17452 | 68 | 0 | 31 | 15 | 2 | 17450 |
| 11 | Peñagrande | 17451 | 69 | 0 | 53 | 14 | 2 | 17450 |
| 12 | Aravaca | 17452 | 68 | 0 | 771 | 71 | 18 | 17434 |
| 13 | Los Cármenes | 17452 | 68 | 0 | 380 | 56 | 8 | 17444 |
| 14 | Opañel | 17452 | 68 | 0 | 101 | 40 | 7 | 17445 |
| 15 | San Diego | 17452 | 68 | 0 | 42 | 26 | 4 | 17448 |
| 16 | Horcajo | 17452 | 68 | 0 | 59 | 12 | 3 | 17449 |
| 17 | Pueblo Nuevo | 17452 | 68 | 0 | 34 | 27 | 5 | 17447 |
| 18 | Canillas | 17452 | 68 | 0 | 124 | 35 | 7 | 17445 |
| 19 | Los Ángeles | 17452 | 68 | 0 | 71 | 25 | 9 | 17443 |
| 20 | Canillejas | 17452 | 68 | 0 | 39 | 11 | 2 | 17450 |
| **AEMET Network** | | | | | | | | |
| 901 | Barajas (reference) | 17516 | 4 | - | - | - | - | - |
| 902 | Retiro | 17428 | 92 | - | - | - | - | - |
| 903 | Cuatro Vientos | 17507 | 13 | - | - | - | - | - |
| 904 | C. Universitaria | 17132 | 388 | - | - | - | - | - |
| 905 | El Goloso | 16983 | 537 | - | - | - | - | - |

*2.4.   Methods*

The relationship between LCZs and intra-urban temperature variability was appraised from two perspectives: statistical and graphical. The latter was used to explore the evolution of the intra-urban and intra-daily temperature differences, mostly in terms of the UHI intensity. In that sense, the UHI intensity was defined as the temperature difference between two LCZs (Stewart and Oke, 2012). As in previous studies (e.g. Budhiraja et al., 2020; Kwok et al., 2019; Skarbit et al., 2017; Yang et al., 2020b), we used a LCZ D site placed on the outskirts of the city as the reference for estimating the UHI intensity in all urban sites (Barajas AEMET Observatory). Therefore, the UHI intensity was estimated as $\Delta T_{LCZ\ X,\ LCZ\ D}$, where the LCZ X represents each of the built-up LCZs. Temperatures at each LCZ were spatially averaged to obtain a representative value for each one (Stewart et al., 2014). Only in the case of the LCZ 9 was the temperature derived from a single measurement site. Regarding the cooling rates, these were estimated as the temperature difference between two consecutive hours, $T_t - T_{t-1}$.

Within the statistical approach, correlation coefficients (Spearman and Pearson, depending on the type of parameter) were used to quantify and numerically compare the LCZ scheme's ability to capture daily, diurnal and nocturnal, temperature disparities. They were also applied to analyse the relevance of each LCZ indicator on an hourly basis. Then, the authors employed Analysis of Variance (ANOVA) models to assess whether the average UHI intensities at each LCZ were statistically different ($p < 0.05$). It should be mentioned that each LCZ had equal-sized samples with a quasi-normal distribution, leading the authors to consider parametric approaches. However, since the Levene's variance check revealed that our LCZs presented different variances for the mean and minimum UHI intensities, a Welch ANOVA was used instead of the classical one-way ANOVA. Together with the Welch ANOVA, the Games-Howell post-hoc test was applied to evaluate if the divergences between each LCZ pair were statistically significant. In this respect, previous studies have also reported the use of non-parametric post-hoc tests, such as the Kruskal-Wallis (Fenner et al., 2017; Leconte et al., 2017) or the Wilcoxon Rank-sum (Chapman et al., 2017) tests. While these present the advantage of not having to comply with any specific distribution, their statistical power might be diminished compared to parametric approaches.

The statistical differences between the LCZs were determined based on days under ideal conditions on an hourly basis. These were defined using the weather factor, $\Phi_w$, introduced by Oke (Oke, 1998, as found in Runnalls & Oke, 2006):

$$\Phi_w = \frac{1 - kn^2}{\sqrt{u}} \qquad \text{where } u \geq 1 \text{ m/s}$$

$\Phi_w$ depends on the wind speed in m/s ($u$) and the cloud cover, which in turn is computed based on the amount of clouds in tenths ($n$) and a correction coefficient depending on the clouds' height ($k$). This way, the lower the cloud cover and the wind speed (i.e. the closer $\Phi_w$ is to 1), the more favourable might weather conditions be. Following the example of previous studies (Skarbit et al., 2017; Stewart et al., 2014; Yang et al., 2018), days were tagged as ideal when $\Phi_w$ > 0.7, which was estimated as a daily average. Since temperature differences are sharper during the nighttime and might depend on the preceding hours' conditions, the daily average was estimated between middays (12:00 UTC).

## 3.  Results

### 3.1.  *Global overview of the UHI of Madrid*

**Table 4** displays the annual averages regarding temperature and UHI intensity at each location during 2017, which was the only full calendar year of the series. In Spain, 2017 was the hottest year since 1965. During this year, all urban sites were mostly warmer than the AEMET Barajas observatory. On average, temperature differences of up to 2.2 ºC can be expected between sites. This variation rises to 4.5 ºC when looking at minimum temperatures and drops to 1.7 ºC for maximum temperatures. Despite not being completed, the 2016 and 2018 series follow the same trend.

A high correlation between the sites registering the highest minimum temperatures and those with the highest maximum UHI intensity was observed. As expected, the UHI presented a clear nighttime pattern, typically registering maximum UHI intensities above 4 ºC. Overall, an Urban Cool Island could be observed as well, although its intensity usually remained below 2 ºC.

**Table 4.** Mean daily temperature and UHI intensity registered for each location during 2017.

| Sensor ID | Site name | LCZ | Dist. to centroid [1] (km) | Temperature (ºC) | | | UHI intensity (ºC) [2] | | | Ranking UHI intensity |
|---|---|---|---|---|---|---|---|---|---|---|
| | | | | Mean | Min | Max | Mean | Min | Max | |
| **MODIFICA Network** | | | | | | | | | | |
| 01 | Embajadores | 2 | 2.0 | **17.9** | **13.3** | 22.5 | **2.2** | -0.7 | **5.2** | **1** |
| 03 | La Chopera | 2 | 3.5 | 17.5 | 12.3 | 22.7 | 1.8 | **-0.6** | 4.3 | 4 |
| 04 | Estrella | 4 | 2.1 | 17.3 | 12.5 | 22.1 | 1.6 | -1.0 | 4.3 | 7 |
| 05 | Pacífico | 2 | 2.3 | 17.6 | 13.1 | 22.3 | 1.9 | -1.0 | 4.9 | 2 |
| 06 | Guindalera | 2 | 1.6 | 17.0 | 12.4 | 22.6 | 1.3 | -1.4 | 4.3 | 9 |
| 07 | Recoletos | 2 | 0.3 | 17.4 | 12.9 | 21.9 | 1.7 | -1.3 | 4.9 | 5 |
| 08 | Hispanoamérica | 2 | 3.2 | 17.0 | 12.4 | 21.8 | 1.3 | -1.6 | 4.4 | 9 |
| 09 | Cuatro Caminos | 2 | 3.6 | 17.4 | 12.9 | 21.9 | 1.7 | -1.3 | 5.0 | 5 |
| 10 | Arapiles | 2 | 2.5 | 17.2 | 13.0 | 21.4 | 1.4 | -1.7 | 4.9 | 8 |
| 11 | Peñagrande | 4 | 6.9 | 16.5 | 11.7 | 22.6 | 0.8 | -1.8 | 3.6 | 15 |
| 12 | Aravaca | 6 | 9.0 | 16.1 | 9.9 | 23.1 | 0.4 | -1.7 | 2.5 | 18 |
| 14 | Opañel | 5 | 5.2 | 17.0 | 11.7 | **23.1** | 1.3 | -1.0 | 3.6 | 9 |
| 15 | San Diego | 2 | 3.8 | 17.6 | 12.7 | 22.6 | 1.9 | -0.6 | 4.7 | 2 |
| 16 | Horcajo | 9 | 4.9 | 16.2 | 11.1 | 21.9 | 0.5 | -1.9 | 3.2 | 17 |
| 17 | Pueblo Nuevo | 2 | 3.4 | 17.0 | 12.1 | 22.4 | 1.3 | -1.1 | 4.0 | 9 |
| 18 | Canillas | 6 | 4.5 | 16.9 | 11.6 | 23.0 | 1.2 | -1.0 | 3.4 | 13 |
| 19 | Los Ángeles | 5 | 7.6 | 16.3 | 11.1 | 22.3 | 0.6 | -1.7 | 3.2 | 16 |
| 20 | Canillejas | 5 | 6.5 | 16.9 | 11.7 | 23.0 | 1.2 | -0.9 | 3.5 | 13 |
| **AEMET Network** | | | | | | | | | | |
| 901 | Barajas (reference) | D | 11.6 | 15.7 | 8.8 | 22.6 | 0.0 | 0.0 | 0.0 | - |

As illustrated in **Fig. 3,** the mean, maximum and minimum UHI intensities (herein after referred to as "UHII$_{mean}$", "UHII$_{max}$", "UHII$_{min}$") also exhibit different behaviour when analysed monthly. Since the UHI is a nighttime phenomenon and nights grow longer close to the winter solstice, it is not surprising to find higher means and medians near the winter solstice, while yielding lower means and medians near the summer solstice. Nevertheless, this trend seems to be clearly influenced by meteorological conditions. For instance, the daily UHII$_{mean}$ in March 2018 was 0 ºC, a far cry from what could be expected for this month, but in line with the aggregation of weather instability shown in **Fig. 4** (bottom). The accumulated precipitation in March 2018 was, in fact, the highest ever recorded since 1965 (AEMET, 2019).

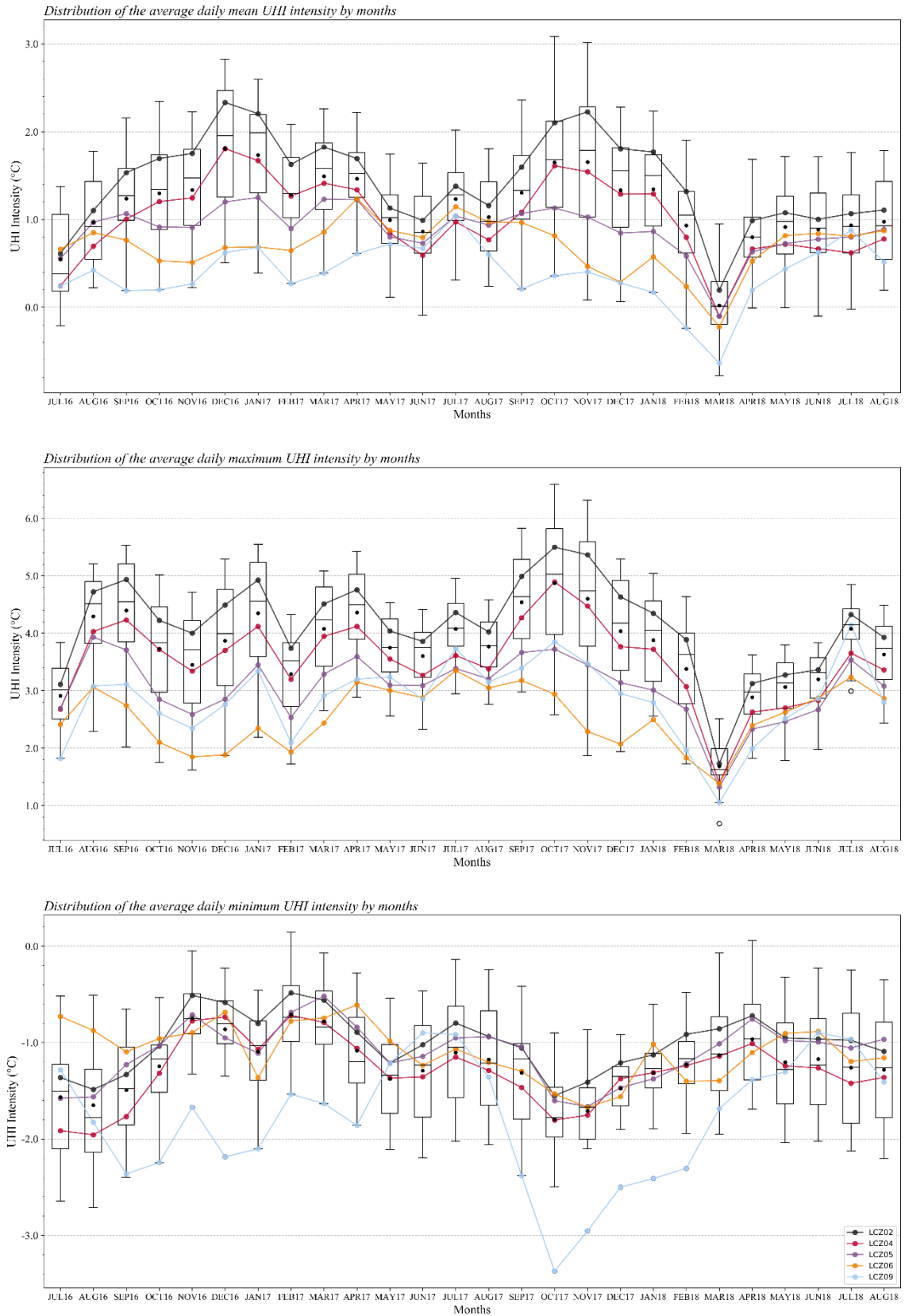

**Fig. 3.** Monthly mean (top), maximum (middle) and minimum (bottom) UHI intensity for all urban sites, between 2016 and 2018. The UHI intensity is expressed as $\Delta T_{LCZ\ X,\ LCZ\ D}$, where LCZ X represents each of the built-up LCZs and LCZ D refers to the AEMET Barajas observatory.

As expected, the daily UHII$_{max}$ also manifested a high sensitivity level to meteorological stability. Yet, it did not seem to follow a seasonal pattern, which contrasts with previous evidence (e.g. Arnds et al., 2017; Fenner et al., 2014; Schatz and Kucharik, 2014). **Fig. 4** reveals that the cluster of days above certain UHI intensities thresholds reflects a homogeneous distribution throughout the year, reaching its highest values mostly with higher atmospheric stability. Nevertheless, it should be underlined that the weather factor ($\Phi_w$) did not seem fully reliable for identifying ideal days. Of all days reaching a $\Phi_w \geq 0.7$ (see **section 2.4**), only 77.7% had temperature differences exceeding 4 ºC and from those, 32.2% remained undetected. It is therefore not surprising that, even though the $\Phi_w$ exhibited a noticeable tendency towards higher values in winter (also found in Yang et al., 2018), the daily UHII$_{max}$ did not replicate this seasonal pattern.

The daily UHII$_{min}$ displayed a different behaviour, with no clear seasonal pattern or discernible effects derived from atmospheric instability. All months showed similar dispersion characteristics, with the exception of the interquartile ranges standing out for appearing larger during the summer months. The latter could be linked to the increase in solar radiation, but this relationship should be further explored.

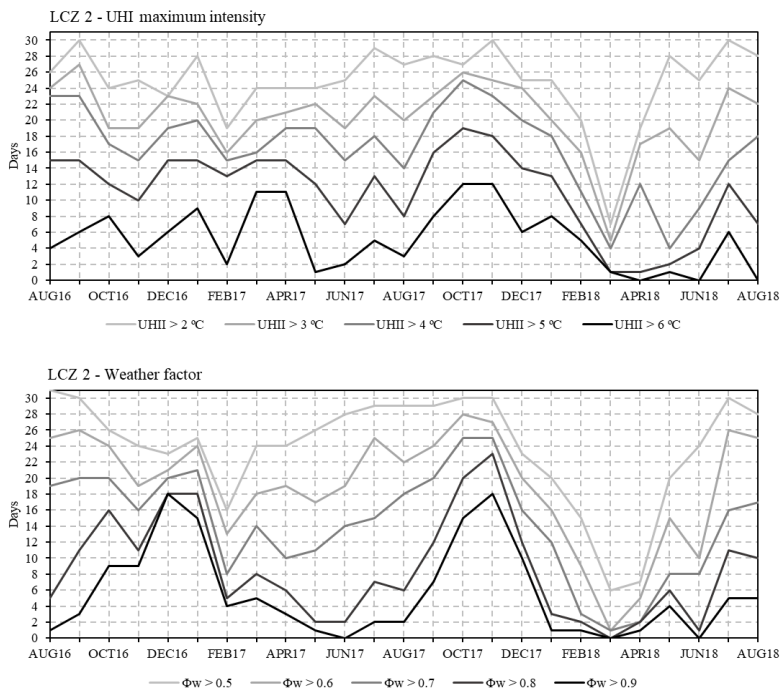

**Fig. 4.** Frequency of days for LCZ 2 in which the UHI maximum intensity (top) and the weather factor (bottom) were above certain values.

## 3.2. *Temperature variability across LCZs on an hourly basis*

The temperature variability was subsequently explored in relation to each LCZ on an hourly basis. The findings from this hourly analysis supported the yearly patterns discussed in the previous section. According to the results presented in **Fig. 5,** temperature differences are sharper during the nighttime regardless of the time of year, thus making the months with longer nights more prone to a higher daily $UHII_{mean}$. The effect of meteorological instability is also evident, as observed in March 2018 (dashed lines). Otherwise, temperature differences appear to be quite similar throughout the year for all LCZs. Once more, no seasonal pattern was observed, in line with the overall trend observed in the previous section.

It is noteworthy that all LCZs share a similar trend, which is undoubtedly related to the traditional UHI profile: temperatures start to greatly differ just after sunset (between 17:00 and 20:00 UTC, depending on the time of year) and reach their maximum disparity just before sunrise (between 5:00 and 8:00 UTC). During the daytime, the UHI seems to fade away and, under certain circumstances, even transform into an Urban Cool Island. It may be noted that, during the nighttime, variations between LCZs become more evident. LCZ 2 and LCZ 6 distinctly and consistently differ from the other LCZs, registering the highest and lowest UHI intensities throughout the year, respectively. LCZs 4 and 5, on the contrary, do not appear to significantly vary from each other: in fact, their UHI show a rather analogous hourly evolution, with differences rarely exceeding 0.2 °C (which concurs with the sensors' accuracy). In theory, LCZ 9 should have been registering the lowest nighttime UHI intensities, yet it matched the values registered for LCZ 4 and 5, above those for LCZ 6. Per contra, LCZ 9 tends to register lower temperatures during the daytime.

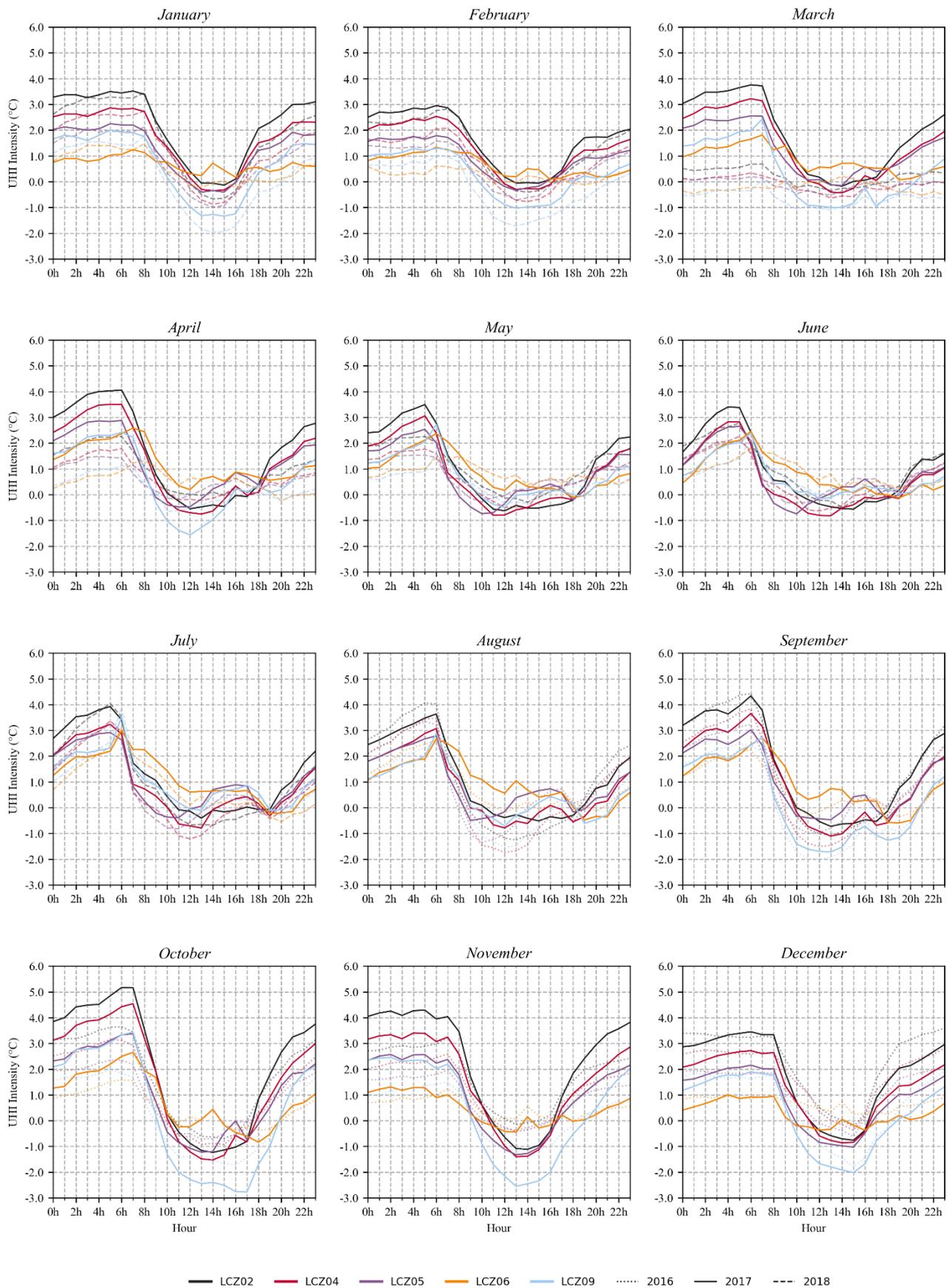

**Fig. 5.** Average hourly UHI intensity from January to December for each LCZ. The data corresponding to the year 2017 are shown in the foreground (solid lines). Measurements corresponding to 2016 and 2018 complete the 2-year series and are displayed in the background (dashed lines). All the data correspond to the average hourly UHI intensity for each of the LCZ included in this study.

To better analyse LCZs particularities, **Fig. 6** provides an in-depth look into two months of the year, presenting the hourly evolution of the temperature, the UHI intensity and the cooling rates across different LCZs. As would have been expected, temperature oscillations are greater during the summer than during the winter months. Stronger oscillations can be found in more open areas, i.e. LCZ 6 and the reference site (LCZ D). LCZ 9 shows a different tendency, with relatively marked oscillations in summer but somewhat weak in winter. No clear explanation is available for this behaviour, apart from representativeness issues associated with the location and source area of the measuring site, since this data came from a single site (#16, Horcajo).

During the daytime, temperature differences between most LCZs are blurred. Although inter-urban thermal differences average nearly 2 ºC in winter (see **Fig. 6** boxplots in the background), the temperature variation range between LCZs 2, 4, 5 and 6 seems to be within 0.2 ºC. A similar pattern is observed in summer, with no clear differentiation between LCZs, despite average temperature variations of up to 2.5 ºC. In summer, however, the average trend of the urban measurement sites unveils a slight overheating of some sensors between 14:00 and 18:00 UTC. This late afternoon overheating gains more prominence in the LCZs that are sparsely built and is especially noticeable close to the summer solstice only to disappear close the winter solstice (see **Fig. 5**). The remaining urban AEMET observatories did not replicate this trend, but the latter is consistent with the findings of previous studies (Fenner et al., 2014; Skarbit et al., 2017; Yang et al., 2020b). A discussion on this topic is included in the next section.

Sharper temperature oscillations also lead to steeper cooling rates (**Fig. 6**, bottom) during the hottest months. In July, the cooling and heating rates are nearly symmetric throughout the day, since each cycle takes about a half-day. In this sense, the cooling cycle starts 2-3 hours before sunset (17:00 UTC) and ends at sunrise (5:00 UTC). In January, however, the cooling cycle takes place during two-thirds of the day, being, therefore, less intense than the heating cycle. It can be observed that during several hours of the night (2:00 – 7:00 UTC) the cooling rate remains very low (<0.2 ºC/h), contributing to a higher UHI nighttime stability during the winter season.

The cooling rate analysis also confirms that a great part of the UHI intensity develops a few hours after sunset (Holmer et al., 2007; Leconte et al., 2017). During the first 3 hours of nighttime, the densest LCZs (02, 04 and 05) reach up to 40-55% of the daily $UHII_{max}$, with significant diversity between LCZs. As shown in **Fig. 6**, these differences then tend to fade away throughout the night. This effect was also observed in previous studies (e.g. Thomas et al., 2014; Yang et al., 2018).

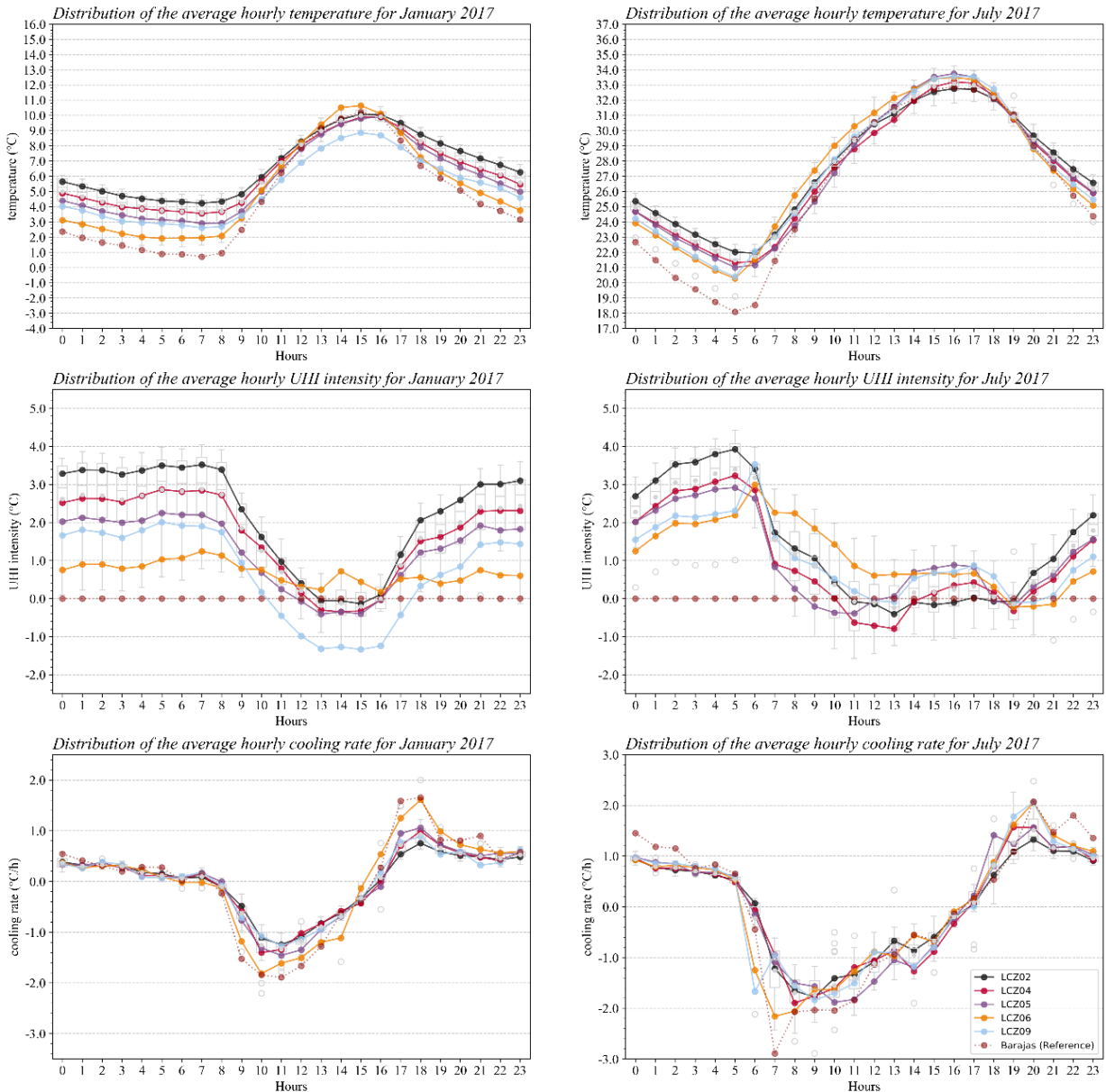

**Fig. 6.** Average hourly temperature (top), UHI intensity (centre) and cooling rates (bottom) for January (left) and July 2017 (right). The different LCZs and the reference site are outlined. The hourly ranges from all measuring sites are displayed as a boxplot in the background.

*3.3. Temperature differences across LCZs from a statistical perspective*

In this section, the authors further investigate the previously identified trends from a statistical perspective. As a starting point, a Spearman's rank correlation coefficient ($r_s$) analysis between the LCZs and the UHI intensity is presented. **Table 5** shows the results for the annual average of the daily $UHII_{mean}$, $UHII_{max}$ and $UHII_{min}$. Sites are ranked according to their LCZ class, from most compact areas (LCZ 2) to most sparsely built areas (LCZ 9). Within the LCZs, the sites were organised according to a random position scheme ($LCZ_{RAND}$), and alternatively via the criteria of distance from the city centre ($LCZ_{DIST}$). The findings reveal that, while there is a strong correlation between the $LCZ_{RAND}$ and the $UHII_{max}$ (nighttime, $r_s = 0.79$), said correlation becomes very weak regarding the $UHII_{min}$ (daytime, $r_s = 0.24$). Including the criteria of distance from the city centre when ranking urban sites ($LCZ_{DIST}$) increases the contrast between night and day ($r_s = 0.85$ and $r_s = 0.19$, respectively). In regard to annual mean data, the LCZs seem to perform poorly at capturing the minimum urban temperature differences that take place during the daytime.

**Table 5.** Spearman's rank correlation coefficient ($r_s$) between the annual average of the daily $UHII_{mean}$, $UHII_{max}$ and $UHII_{min}$, and the different classification methods. The LCZs were ranked from 1 to 10, and then from A to E. The sites classified within the same LCZ were attributed a random position ($LCZ_{RAND}$) or according to their distance to the urban centre ($LCZ_{DIST}$).

| Class. method | $UHII_{mean}$ | $UHII_{max}$ | $UHII_{min}$ |
|---|---|---|---|
| RAND classification | -0.03 | 0.00 | 0.02 |
| DIST classification | 0.72 | 0.79 | 0.25 |
| $LCZ_{RAND}$ classification | 0.75 | 0.79 | 0.24 |
| $LCZ_{DIST}$ classification | 0.75 | 0.85 | 0.19 |

The Welch ANOVA models and post-hoc Games-Howell tests confirmed these results, revealing varying significant differences between the LCZs in reference to the $UHII_{mean}$, $UHII_{max}$, and the $UHII_{min}$ (**Table 6**). In that sense, LCZ 2 differs from all other LCZs in relation to both the $UHII_{max}$ and $UHII_{mean}$, but it does not for the $UHII_{min}$. LCZ 4 and 5 always group together, which is consistent with the trends observed in **Fig. 5**. LCZ 4 and 5 might also associate with other LCZ individually, depending on the moment of the day. LCZ 6 and 9 also group together in terms of the $UHII_{mean}$ and $UHII_{min}$, but they diverge as far as the $UHII_{max}$ is concerned.

Daytime differences found in LCZs should be, however, interpreted with caution. First, because most concentrate in the LCZ 9 class, which is characterised by a single measuring site and which did not conform to the expected behaviour associated with this type of LCZ class during the graphic analysis. And second, due to the high temperature variability during the daytime hours. Situations such as the late afternoon overheating identified in the previous section, where it is unclear if it is intrinsic to the LCZs characteristics, could be influencing these results.

**Table 6.** Games-Howell post-hoc test between pairs of LCZs (significance level α = 0.05). Tested for the average monthly $UHII_{mean}$, $UHII_{max}$, and $UHII_{min}$.

| LCZ pairs | Significant difference between pairs | | |
|---|---|---|---|
|  | $UHII_{mean}$ | $UHII_{max}$ | $UHII_{min}$ |
| 2 - 4 | ■ | ■ |  |
| 2 - 5 | ■ | ■ |  |
| 2 - 6 | ■ | ■ | ■ |
| 2 - 9 | ■ | ■ | ■ |
| 4 - 5 |  |  |  |
| 4 - 6 | ■ | ■ |  |
| 4 - 9 | ■ | ■ | ■ |
| 5 - 6 | ■ | ■ | ■ |
| 5 - 9 | ■ |  | ■ |
| 6 - 9 |  | ■ | ■ |

This point is confirmed when looking at the ANOVAs and post-hoc tests on an hourly basis. The results are given for the months of January and July 2017 (see **Table 7**), for every day of the month and the days with ideal conditions. During the daytime, there are more pairs of LCZs associated with LCZ 9. In July, significant differences are also found between other LCZs from 13:00 to 17:00 UTC, which coincides to a large extent with the late afternoon overheating hours. Contrariwise, the results suggest a much higher level of consistency and stability in the differences between LCZs during the nighttime. In January, these differences are more relevant between 1:00 and 3:00 UTC (middle of the night), while in July they intensify between 3:00 and 5:00 UTC (end of the night). These hour bands occur in both cases 7 to 8 hours after sunset.

**Table 7.** Hourly Games-Howell post-hoc test between pairs of LCZs for January and July 2017. This analysis is presented for every day of each month and for the days with ideal conditions ($\Phi_w > 0.7$, no precipitations in the last 24 hours and an UHI intensity > 5 °C). Notice how, within ideal days, significant differences concentrated at different times of the night, depending on the time of the year (dashed square).

| LCZ pairs | 0h | 1h | 2h | 3h | 4h | 5h | 6h | 7h | 8h | 9h | 10h | 11h | 12h | 13h | 14h | 15h | 16h | 17h | 18h | 19h | 20h | 21h | 22h | 23h |
|---|---|---|---|---|---|---|---|---|---|---|---|---|---|---|---|---|---|---|---|---|---|---|---|---|
| **EVERY DAY** | | | | | | | | | | | | | | | | | | | | | | | | |
| **January 2017 (n = 31 days)** | | | | | | | | | | | | | | | | | | | | | | | | |
| 2 - 4 | | | | | | | | | | | | | | | | | | | | | | | | |
| 2 - 5 | ■ | | | | ■ | ■ | ■ | ■ | ■ | | | | | | | | | | | ■ | ■ | ■ | ■ | ■ |
| 2 - 6 | ■ | ■ | ■ | ■ | ■ | ■ | ■ | ■ | ■ | | | | | | | | | | ■ | ■ | ■ | ■ | ■ | ■ |
| 2 - 9 | ■ | ■ | | ■ | ■ | ■ | ■ | ■ | ■ | ■ | | | | | ■ | ■ | ■ | ■ | ■ | ■ | ■ | ■ | ■ | ■ |
| 4 - 5 | | | | | | | | | | | | | | | | | | | | | | | | |
| 4 - 6 | ■ | | | | ■ | ■ | ■ | ■ | ■ | | | | | | | | | | ■ | ■ | ■ | ■ | ■ | ■ |
| 4 - 9 | | | | | | | | | | ■ | ■ | ■ | ■ | ■ | ■ | ■ | ■ | | | | | | | |
| 5 - 6 | | | | | | | | | | | | | | | | | | ■ | ■ | | | | | |
| 5 - 9 | | | | | | | | | | | | ■ | ■ | ■ | ■ | ■ | ■ | | | | | | | |
| 6 - 9 | | | | | | | | | | | ■ | ■ | ■ | ■ | ■ | ■ | | | | | | | | |
| **July 2017 (n = 31 days)** | | | | | | | | | | | | | | | | | | | | | | | | |
| 2 - 4 | | | | | | | | ■ | ■ | ■ | | ■ | ■ | | ■ | | | | | | | | | |
| 2 - 5 | | | ■ | | | ■ | | ■ | ■ | | | | ■ | ■ | | ■ | | | | | | | | |
| 2 - 6 | ■ | ■ | ■ | ■ | ■ | ■ | | | ■ | ■ | | ■ | | | ■ | ■ | | | ■ | ■ | | ■ | ■ | ■ |
| 2 - 9 | ■ | ■ | ■ | ■ | ■ | ■ | | | | | | | | | ■ | ■ | ■ | | ■ | ■ | | ■ | ■ | |
| 4 - 5 | | | | | | | | | | | | | ■ | | | ■ | ■ | ■ | | | | | | |
| 4 - 6 | ■ | ■ | ■ | ■ | ■ | ■ | | | | | | | | | | | | | | | ■ | ■ | ■ | ■ |
| 4 - 9 | | | | | | | | | | | | ■ | ■ | ■ | | | ■ | ■ | ■ | | | | | |
| 5 - 6 | ■ | ■ | ■ | ■ | ■ | ■ | | | | | | | | | ■ | ■ | ■ | | ■ | ■ | ■ | ■ | ■ | ■ |
| 5 - 9 | | | | | | | | ■ | ■ | ■ | ■ | ■ | | | | | | ■ | | | | | | |
| 6 - 9 | | | | | | | | ■ | ■ | ■ | | ■ | ■ | ■ | ■ | ■ | ■ | ■ | ■ | | | | | |
| **IDEAL DAYS** | | | | | | | | | | | | | | | | | | | | | | | | |
| **January 2017 (n = 12-13 days)** | | | | | | | | | | | | | | | | | | | | | | | | |
| 2 - 4 | ■ | ■ | ■ | ■ | ■ | ■ | | | | | | | | | | | | | | | | | | |
| 2 - 5 | ■ | ■ | ■ | ■ | ■ | ■ | ■ | ■ | ■ | ■ | | | | | | | | | | ■ | ■ | ■ | ■ | ■ |
| 2 - 6 | ■ | ■ | ■ | ■ | ■ | ■ | ■ | ■ | ■ | ■ | | | | | | | | | | ■ | ■ | ■ | ■ | ■ |
| 2 - 9 | ■ | ■ | ■ | ■ | ■ | ■ | ■ | ■ | ■ | ■ | ■ | ■ | ■ | ■ | ■ | ■ | ■ | ■ | ■ | ■ | ■ | ■ | ■ | ■ |
| 4 - 5 | | ■ | ■ | ■ | ■ | | | | | | | | | | | | | | | | | | | |
| 4 - 6 | ■ | ■ | ■ | ■ | ■ | ■ | ■ | ■ | ■ | | | | | | | | ■ | | ■ | | ■ | ■ | ■ | ■ |
| 4 - 9 | | ■ | ■ | ■ | | | | | | | | ■ | ■ | ■ | ■ | ■ | ■ | ■ | | | | | | |
| 5 - 6 | ■ | ■ | ■ | ■ | | | | | | | | | | | | | ■ | | | ■ | | | | ■ |
| 5 - 9 | | | | | | | | | | | | | | ■ | ■ | ■ | ■ | ■ | | | | | | |
| 6 - 9 | ■ | ■ | ■ | ■ | ■ | | | | | | | | | ■ | ■ | ■ | ■ | ■ | | | | | | |
| **July 2017 (n = 10 days)** | | | | | | | | | | | | | | | | | | | | | | | | |
| 2 - 4 | | ■ | | | ■ | | ■ | | | | | | | ■ | ■ | | | | | | | | | |
| 2 - 5 | ■ | | ■ | | ■ | ■ | ■ | ■ | ■ | ■ | | | ■ | ■ | ■ | ■ | ■ | | | | | | | |
| 2 - 6 | ■ | ■ | ■ | ■ | ■ | ■ | ■ | | | | | | | | ■ | | | | ■ | ■ | ■ | | ■ | ■ |
| 2 - 9 | ■ | | ■ | ■ | ■ | ■ | | | | | | | | | | | ■ | ■ | ■ | ■ | ■ | | | ■ |
| 4 - 5 | | | | | | | | | | | | | | ■ | ■ | ■ | | | | | | | | |
| 4 - 6 | ■ | ■ | ■ | ■ | ■ | ■ | | | | | | | | | ■ | | | | | | | | | ■ |
| 4 - 9 | | | | | ■ | | | | | | | | ■ | ■ | ■ | ■ | | ■ | | | | | | |
| 5 - 6 | ■ | ■ | ■ | ■ | ■ | ■ | | | | | | | | ■ | ■ | | ■ | | ■ | | | ■ | | ■ |
| 5 - 9 | | | | | | | | ■ | ■ | ■ | ■ | | | | | | | ■ | | | | | | |
| 6 - 9 | | ■ | | ■ | | ■ | ■ | | | | | ■ | | | | ■ | ■ | ■ | ■ | ■ | ■ | | | ■ |

The results show that the LCZs' ability to illustrate differences in urban temperature varies throughout the day. This is examined in further depth via a correlation analysis between LCZ indicators and the UHI intensity. As shown in **Table 8**, none of the LCZ parameters are able to account for temperature differences during the daytime. While most of these parameters strongly correlate with nighttime temperature differences (i.e. previous surface fraction (PSF), aspect ratio (AR), building surface fraction (BSF) and sky view factor in winter (SVFW) show $r_p > \pm 0.7$), these correlations steadily weaken after sunrise until fading away. The impervious surface fraction indicator (ISF) stands out as an exception, linking with daytime urban temperatures only and exhibiting a relatively weak correlation ($r_p < -0.5$).

Most LCZ parameters showed similar correlations during summer and winter, except for the SVF. The annual cycle of deciduous trees, which are prevailing in the measuring sites, is somewhat detected by the SVFS and SVFW. Surprisingly, the SVFW showed slightly better performance during the summer nights, while this was true for the SVFS during the winter daytime.

While it is remarkable that all LCZs presented relevant correlations with the UHI at a certain point in time ($r_p > \pm 0.3$), it might be argued that additional indicators could be included into the scheme for better capturing the urban temperature variability, particularly during the daytime. One of the most discussed additional parameters is the distance to the city centre (DIST, e.g. Gardes et al., 2020; Kotharkar et al., 2019). As shown in **Table 8**, DIST follows the pattern of other LCZ indicators, with strong correlations during the nighttime that lessen during the daytime. In spite of failing to solve the issue of predicting of daytime temperature differences, its predictive power might be on the same level than that of the PSF, which discloses the highest correlations among LCZ parameters ($r_p > -0.8$). Moreover, one might consider that DIST shows strong linear relationships with other LCZs. For example, the PSF tends to be more prominent in the outskirts of the city. It is certainly true that DIST displays a relevant collinearity degree with most LCZ parameters ($r_p > \pm 0.5$, see **Fig. 7**), reaching its maximum with the PSF ($r_p = 0.74$). Nonetheless, its collinearity is akin to that found between LCZ parameters. In fact, the highest values are found between BSF-PSF ($r_p = -0.94$), AR-PSF ($r_p = -0.76$) and AR-SVFW ($r_p = -0.75$).

**Table 8.** Pearson's correlation coefficient between the hourly UHII$_{mean}$ and six LCZ indicators for January and July of 2017.

| | Correlation coefficient ($r_p$) | | | | | | | | | | | | | | | | | | | | | | | |
|---|---|---|---|---|---|---|---|---|---|---|---|---|---|---|---|---|---|---|---|---|---|---|---|---|
| LCZ indicator | 0 h | 1 h | 2 h | 3 h | 4 h | 5 h | 6 h | 7 h | 8 h | 9 h | 10 h | 11 h | 12 h | 13 h | 14 h | 15 h | 16 h | 17 h | 18 h | 19 h | 20 h | 21 h | 22 h | 23 h |
| **January 2017** | | | | | | | | Nighttime | | Daytime | | | | | | | Daytime | Nighttime | | | | | | |
| SVFS | -0.43 | -0.42 | -0.42 | -0.40 | -0.39 | -0.37 | -0.41 | -0.45 | -0.46 | -0.48 | -0.58 | -0.52 | -0.42 | -0.35 | -0.49 | -0.49 | -0.39 | -0.61 | -0.63 | -0.58 | -0.56 | -0.52 | -0.49 | -0.49 |
| SVFW | -0.77 | -0.75 | -0.76 | -0.75 | -0.75 | -0.74 | -0.76 | -0.77 | -0.78 | -0.76 | -0.70 | -0.58 | -0.46 | -0.24 | -0.11 | -0.18 | -0.29 | -0.67 | -0.79 | -0.80 | -0.80 | -0.79 | -0.79 | -0.79 |
| AR | 0.77 | 0.75 | 0.75 | 0.74 | 0.73 | 0.73 | 0.74 | 0.77 | 0.78 | 0.79 | 0.73 | 0.64 | 0.60 | 0.43 | 0.32 | 0.38 | 0.54 | 0.81 | 0.86 | 0.85 | 0.84 | 0.82 | 0.81 | 0.81 |
| BSF | 0.74 | 0.72 | 0.72 | 0.73 | 0.73 | 0.72 | 0.72 | 0.75 | 0.76 | 0.76 | 0.69 | 0.60 | 0.56 | 0.42 | 0.25 | 0.27 | 0.35 | 0.67 | 0.75 | 0.77 | 0.76 | 0.74 | 0.74 | 0.76 |
| ISF | -0.02 | -0.02 | -0.03 | -0.04 | -0.05 | -0.03 | -0.02 | -0.08 | -0.11 | -0.17 | -0.25 | -0.31 | -0.45 | -0.50 | -0.44 | -0.38 | -0.29 | -0.18 | -0.05 | -0.02 | -0.01 | 0.03 | 0.00 | -0.02 |
| PSF | -0.85 | -0.83 | -0.82 | -0.82 | -0.83 | -0.82 | -0.82 | -0.84 | -0.84 | -0.82 | -0.70 | -0.57 | -0.47 | -0.29 | -0.12 | -0.17 | -0.29 | -0.71 | -0.85 | -0.88 | -0.88 | -0.87 | -0.86 | -0.87 |
| HRE | 0.27 | 0.27 | 0.26 | 0.26 | 0.28 | 0.29 | 0.29 | 0.28 | 0.28 | 0.25 | 0.25 | 0.16 | -0.06 | -0.18 | -0.29 | -0.20 | -0.01 | 0.15 | 0.25 | 0.27 | 0.28 | 0.30 | 0.31 | 0.28 |
| DIST | -0.85 | -0.84 | -0.84 | -0.82 | -0.82 | -0.82 | -0.84 | -0.85 | -0.85 | -0.84 | -0.66 | -0.51 | -0.32 | -0.12 | -0.06 | -0.09 | -0.13 | -0.59 | -0.79 | -0.83 | -0.86 | -0.88 | -0.88 | -0.87 |
| **July 2017** | | | | | | | | | | | | | | | | | | | | | | | | |
| SVFS | -0.63 | -0.65 | -0.67 | -0.67 | -0.69 | -0.69 | -0.51 | -0.28 | -0.45 | -0.58 | -0.54 | -0.46 | -0.40 | -0.10 | 0.17 | 0.20 | 0.24 | 0.30 | -0.02 | -0.34 | -0.58 | -0.61 | -0.63 | -0.64 |
| SVFW | -0.75 | -0.77 | -0.80 | -0.80 | -0.81 | -0.81 | -0.53 | -0.09 | -0.16 | -0.24 | -0.19 | -0.05 | 0.02 | 0.25 | 0.36 | 0.33 | 0.36 | 0.38 | 0.11 | -0.26 | -0.60 | -0.66 | -0.68 | -0.73 |
| AR | 0.76 | 0.77 | 0.79 | 0.80 | 0.81 | 0.82 | 0.52 | 0.09 | 0.15 | 0.29 | 0.26 | 0.21 | 0.15 | -0.07 | -0.22 | -0.25 | -0.26 | -0.43 | -0.31 | 0.25 | 0.65 | 0.70 | 0.72 | 0.74 |
| BSF | 0.75 | 0.78 | 0.77 | 0.76 | 0.76 | 0.77 | 0.64 | 0.38 | 0.26 | 0.32 | 0.13 | 0.12 | 0.12 | -0.05 | -0.25 | -0.37 | -0.41 | -0.48 | -0.19 | 0.29 | 0.60 | 0.64 | 0.68 | 0.71 |
| ISF | -0.05 | -0.07 | -0.07 | -0.04 | -0.03 | -0.06 | -0.40 | -0.64 | -0.50 | -0.46 | -0.24 | -0.25 | -0.22 | -0.16 | 0.09 | 0.18 | 0.28 | 0.24 | 0.01 | -0.06 | 0.00 | 0.03 | 0.00 | 0.01 |
| PSF | -0.85 | -0.87 | -0.87 | -0.87 | -0.86 | -0.86 | -0.58 | -0.19 | -0.11 | -0.19 | -0.06 | -0.04 | -0.05 | 0.12 | 0.25 | 0.36 | 0.36 | 0.47 | 0.21 | -0.31 | -0.69 | -0.75 | -0.79 | -0.83 |
| HRE | 0.09 | 0.11 | 0.14 | 0.15 | 0.15 | 0.17 | -0.03 | -0.36 | -0.19 | -0.06 | -0.02 | -0.15 | -0.25 | -0.41 | -0.33 | -0.25 | -0.17 | -0.28 | -0.20 | -0.11 | 0.03 | 0.08 | 0.08 | 0.11 |
| DIST | -0.78 | -0.80 | -0.82 | -0.82 | -0.82 | -0.82 | -0.67 | -0.33 | -0.34 | -0.39 | -0.22 | -0.11 | -0.01 | 0.19 | 0.43 | 0.51 | 0.50 | 0.63 | 0.32 | -0.21 | -0.59 | -0.67 | -0.71 | -0.75 |
| | Nighttime | | | | | | Daytime | | | | | | | | | | | | Daytime | Nighttime | | | | |

|      | DIST  | SVFS  | SVFW  | AR    | BSF   | ISF   | PSF  | HRE  |
|------|-------|-------|-------|-------|-------|-------|------|------|
| DIST | 1.00  |       |       |       |       |       |      |      |
| SVFS | 0.57  | 1.00  |       |       |       |       |      |      |
| SVFW | 0.64  | 0.71  | 1.00  |       |       |       |      |      |
| AR   | -0.69 | -0.57 | -0.75 | 1.00  |       |       |      |      |
| BSF  | -0.66 | -0.35 | -0.57 | 0.73  | 1.00  |       |      |      |
| ISF  | 0.07  | -0.05 | -0.08 | -0.21 | -0.55 | 1.00  |      |      |
| PSF  | 0.74  | 0.42  | 0.69  | -0.76 | -0.94 | 0.24  | 1.00 |      |
| HRE  | -0.28 | -0.16 | -0.23 | 0.13  | -0.24 | 0.60  | 0.04 | 1.00 |

**Fig. 7.** Correlation matrix between the following LCZ parameters: sky view factor in summer (SVFS) and winter (SVFW), aspect ratio (AR), building surface fraction (BSF), impervious surface fraction (ISF), pervious surface fraction (PSF) and height of roughness elements (HRE). It also includes the distance to the city centre (DIST).

## 4. Discussion

The results presented in this paper support previous evidence suggesting that LCZs can correctly capture most of the intra-urban nighttime temperature variability. The hourly behaviour and expected cooling rates associated with each LCZ, which are primarily defined by the nighttime pattern of the UHI, are also in consonance with previous findings. However, LCZs fail to discern the daytime temperature variability to a large extent. LCZ indicators also exhibited much better correlations with temperature differences during the nighttime than during the daytime, with none providing a good estimate during the central hours of the day. There is no doubt that urban temperature differences are considerably wider during the nighttime than during the daytime, but the latter are still significant (>2 ºC). It would also be interesting to consider whether diurnal temperatures could be framed within a classification system like the LCZs, or if they are heavily dependent on the micro-climatic characteristics of urban areas. In either case, and regardless of the higher uncertainties introduced by solar radiation and the varying urban geometry, further research is warranted to better comprehend the determining factors behind the daytime thermal differences.

A limitation of this study lies in the fact that it only focuses on six out of ten LCZ parameters. The other four (the anthropogenic heat output, the terrain roughness, the surface albedo, and the surface admittance) were excluded due to the lack of robust data. Although these could be approximated based on the metadata included in the Appendix, these estimations would be subject to interpretation, which could include too many uncertainties. It could also be argued that these parameters might be similar across all the LCZs within the same city, or that they might strongly vary within each LCZ. In any case, the absence of some LCZ parameters is a common occurrence in most previous investigations and has not constituted an impediment to define and study the LCZs (Kwok et al., 2019; Yang et al., 2018). Furthermore, the high correlation between some of these indicators is also notable (Kotharkar et al., 2019; Leconte et al., 2020), removing the need to use them in their entirety for classifying urban areas into LCZs. The WUDAPT Project is an example of this procedure, in which a random forest classifier algorithm is trained to classify urban areas into LCZs based only in 2-D multispectral satellite imagery (Bechtel et al., 2019, 2015). When analysing their differences at the intra-urban and hourly level, however, including these other parameters might contribute to gaining a better insight into the climatic properties of each LCZ. In fact, they might be particularly relevant for representing diurnal temperature variations, since albedo and surface admittance are closely linked to heat storage from solar radiation.

The inclusion of the distance to the city centre (DIST) as an extra indicator within the LCZ scheme should also be discussed. This research has demonstrated that, when compared with the other LCZs parameters, DIST ranked among those with the highest correlation coefficients. Several previous studies have also emphasised its relevance for examining urban temperature differences (e.g. Gardes et al., 2020; Kotharkar et al., 2019; Kwok et al., 2019; Leconte et al., 2020, 2017, 2015). This might be particularly relevant during calm nights with clear skies, when the air temperature differences between the urban areas and their surroundings promote the entrance of cool air into the city (Oke et al., 2017b). This *country breeze* would cool down the outskirts of the urban areas, but its effect might progressively abate as we get close to the city centre. In like manner, this cooling effect can be observed at the intra-urban level with contrasting LCZs (e.g. green areas next to densely built-up areas), albeit in a localised and avoidable fashion by virtue of proper sensor siting. On the other hand, this same breeze might result in a rural-to-urban thermal gradient (i.e. from the outskirts to the city centre), affecting urban areas regardless of their LCZ class.

It should also be mentioned that this airflow behaves to a large extent independently of where the city is established, and should, therefore, be differentiated from other wind flows that only originate when the city is located in a specific place, such as the sea or valley breezes. The fact that the countryside breeze is an urban intrinsic phenomenon might be a determining factor for its inclusion as a parameter to classify and compare the climatic properties of an urban area. What is more, it might call into question the comparability of two built-up sections, classified in the same LCZ but positioned differently within an urban area. In addition, it could also help overcome the partial overlap between LCZs found in previous studies (Fenner et al., 2017). To this respect, it would be worth investigating if the need to divide LCZs into subclasses could be reduced when taking DIST into account, and how this could be extended to non-concentric urban areas.

In the present study, only a subtle seasonal pattern was noticed during winter for the $UHII_{mean}$, which could be linked to the night length during this period. Unlike previous studies, no clear seasonal pattern was identified for the $UHII_{max}$. It is worth mentioning that previous evidence suggests that the most severe temperature differences might appear during the warmer months of the year, as happens in Madison (Schatz and Kucharik, 2014), Szeged (Skarbit et al., 2017), Berlin (Fenner et al., 2017, 2014), Hamburg (Arnds et al., 2017) or Lodz (Klysik and Fortuniak, 1999). However, other studies also point to pronounced UHI intensities during the winter (e.g. Glasgow, (Emmanuel and Krüger, 2012), Beijing (Yang et al., 2013), Nanjing (Yang et al., 2020a) or Buenos Aires (Figuerola and Mazzeo, 1998)). In the case of Madrid, previous investigations suggest that the widest thermal differences take place both in the winter (Fernández García et al., 2016) and the summer periods (Núñez Peiró et al., 2017; Yagüe et al., 1991).

The reasons for these seasonal discrepancies remain unclear. Multiple factors have been singled out for promoting seasonality. For example, some studies have related this occurrence to parameters presenting a clear annual cycle, such as solar irradiation (Arnds et al., 2017; Núñez Peiró et al., 2017) or the foliation and defoliation tree cycle (Stewart et al., 2014; Yang et al., 2018). Other works point towards the accumulation of favourable meteorological conditions during a specific time of the year, particularly wind, cloudiness, and precipitation (Yang et al., 2020a). This is evident in subtropical areas affected by the monsoon (Thomas et al., 2014), as well as in mid-latitude cities in which certain meteorological conditions might concentrate on specific months (e.g. Berlin, Fenner et al., 2017). Our measurements outline the relevance of certain meteorological conditions in the formation of the UHI (e.g. March 2018, section 3.1), but these conditions do not seem to accumulate during any particular season. The concentration of ideal days with UHI intensities > 5 ºC was, in fact, slightly larger during January than during July 2017 (13 vs 10 days, respectively), and a similar pattern was found during the rest of the two-year measurement period. To this respect, although the meteorological instability in Madrid tends to aggravate during the spring and autumn, the cluster of unfavourable conditions in March 2018 seems more likely to result from transitory meteorological circumstances than from the seasonal climatic background of the city.

The accuracy level of the collected data concerning diurnal thermal disparities among LCZs is also worth discussing. As expected, temperatures within urban areas were mostly equal or below those registered at the reference site during the daytime, and significantly lower than during the nighttime. However, temperature differences at the more sparsely built areas (i.e. LCZ 4, 5, 6 and 9) suddenly swapped to positive values during the late afternoon of several months (12-18 GMT) only to turn back to negative values a few hours before sunset, when the nighttime UHI starts to form. This late afternoon overheating at urban sites exacerbates when approaching the summer solstice, when solar radiation is at its strongest and the sun is at its maximum altitude. A visual analysis of the data reported in previous investigations revealed similar trends (Fenner et al., 2014; Skarbit et al., 2017; Yang et al., 2020b). Erell & Williamson (2007) determined that urban-rural daytime temperature differences are heavily connected with solar radiation exposure. They also proved that significant thermal differences between urban sites should be expected due to the orientation of streets during the afternoon and late afternoon. Fenner et al (2014) compared downward short-wave radiation within an urban canyon and on the rooftop, connecting the increase in temperature registered during the afternoon at the two urban sites with the orientation of the streets (N-S).

However, one might debate whether this daytime temperature anomaly is exclusively attributed to higher solar access in view of street orientation, or if it could be related to sensor overheating. The majority of the previous research on LCZs has used Stevenson-like, naturally-ventilated radiation shields (Beck et al., 2018; Fenner et al., 2014; Kotharkar and Bagade, 2018; Yang et al., 2018), possibly compromising ventilation when exposed to high levels of solar radiation. In this context, Erell et al. (2005) observed temperature differences of around +1 ºC when comparing a Stevenson screen and a mechanically ventilated shield in the late afternoon of a sunny day. In like manner, the tests of Fenner et al. (2014), who checked their daytime urban measurements against a sensor housed within a mechanically ventilated radiation shield, found thermal differences of up to 0.4 (±0.5) ºC. Although our equipment was designed to be ventilated mechanically and its correct functioning was continuously ensured during the measurement campaign, a similar overheating phenomenon took place. Consequently, studies in cities with high-intensity levels of solar radiation during the summertime, like Madrid, might require extra precautions to avoid sensor overheating. These observations might also put into question the daytime LCZ statistical analysis, since the significance found between LCZ pairs might be due not to the intrinsic climatic properties of each LCZ, but to the fact that some of them might provide enhanced solar radiation access and, therefore, account for stronger sensor overheating episodes.

Finally, it should be mentioned that it was surprising to find overheating sensors despite following the QC procedures derived from the WMO recommendations. It should be noted that other urban studies have made use of advanced filtering techniques to remove or correct their measurements, particularly those based on CWS, where the correct siting, housing and calibration of the sensor cannot be guaranteed (Chapman et al., 2017; Hammerberg et al., 2018; Meier et al., 2017). Others, such as Beck et al. (2018), designed spatial consistency tests based on reference time series derived from inverse distance weighting (IDW) spatial interpolations. Given that the WMO already provides recommendations for sensor siting and metadata collection within urban environments, it would be desirable to include specific guidelines on QC within urban areas, particularly considering the use of non-standard equipment. When using spatial consistency tests, it would be advisable to work with an official reference site, since using averages from the deployed sensors could mask overheating issues.

## 5. Conclusions

This study allowed for a deeper understanding of the UHI of Madrid, as well as the capacity of LCZs to portray these temperature differences within the urban environment. In relation to the former, relatively high UHI intensities were consistently observed in Madrid, reaching values above 5 ºC in more than 35% of the total monitoring days. There was no clear seasonal pattern found for the maximum and minimum UHI intensities, while a slight seasonality towards higher daily means could be noted in winter.

For their part, the LCZs performed strongly concerning the detection of the UHI nighttime profile. They showed high levels of correspondence with the logical LCZs arrangement, confirming that compact urban settings (LCZ 2) systematically register higher UHI intensities than sparsely built ones (LCZ 6). During the night of ideal days, statistical differences proved to be significant among all the LCZs as well, particularly 7 to 8 hours after sunset. During the day, however, the LCZs did not seem to be very effective. The correlation study between the LCZ parameters and the temperature differences corroborates this point, with all parameters failing to accurately represent midday temperature differences. Although diurnal temperature divergences might be closely linked with micro-climatic parameters, such as the street orientation or the sensor location within the urban canyon, more research is needed to identify relevant drivers of diurnal temperature differences and their possible incorporation into the LCZ classification scheme. It would also be advisable to further explore the role of the distance to the city centre within the LCZs.

Overall, the instruments used for the monitoring campaign, as well as the tools adopted for the contextualisation and pre-processing, have shown a high level of reliability. Doubts have only arisen with temperature records during the central hours of the days with the strongest level of solar radiation. It is unclear whether the so-called late afternoon overheating was due to actual sensor overheating issues or to the increased solar radiation availability in the urban canyon. In any case, it would be advisable to continue investigating new ways of protecting the measuring devices as well as novel QC processes to detect anomalies within urban environments.

# 6. Appendix

*Physical characteristics of the MODIFICA measuring campaign*
20 measuring units (sensor + radiation shield)
Global dimensions of each unit: 25 cm x 25 cm x 28 cm.
Weight: 1.1 Kg.
Equipped with an exhaust fan and a photovoltaic panel. It does not require an electrical connection.

*Location*
Fixed on street lampposts.
Height: ~ 6 meters
Mechanically fixed with steel clamps.

*Length of the measuring campaign*
Start date: July 22$^{nd}$, 2016.
Finish date: October 14$^{th}$, 2019.
1$^{st}$ inspection: 1 month after deployment.
Regular inspections every 6 months.

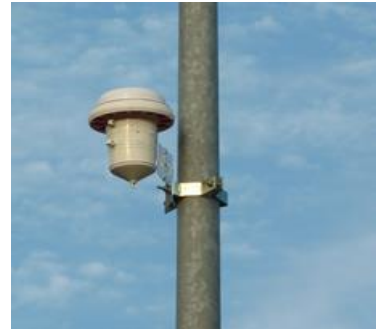

During each inspection, data was downloaded, the equipment was cleaned, and its overall correct functioning was assessed, including the activation of the exhaust fan.

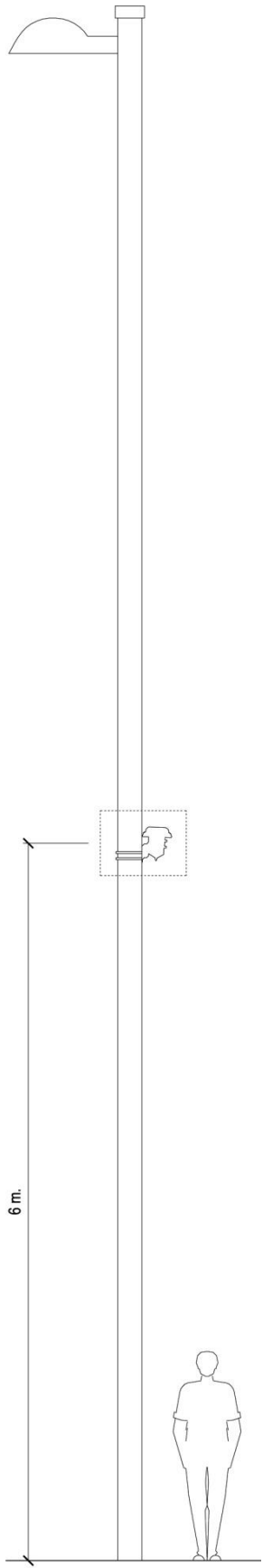

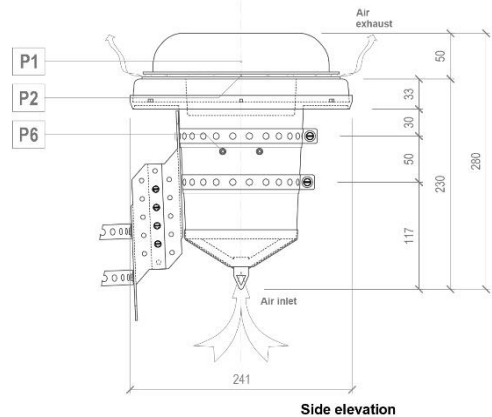
Side elevation

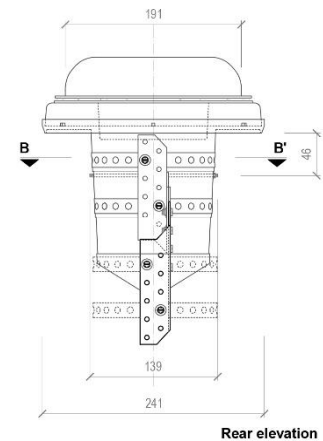
Rear elevation

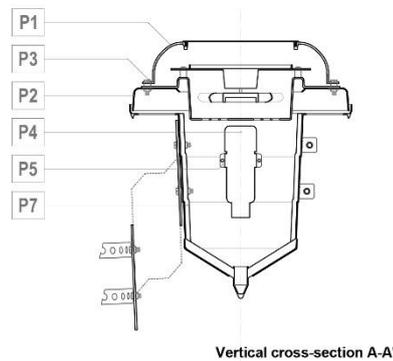
Vertical cross-section A-A'

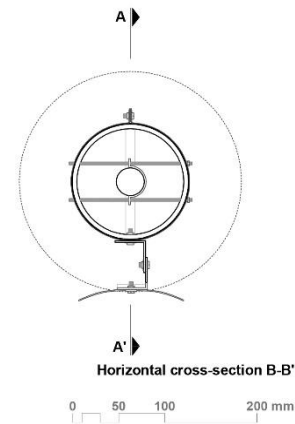
Horizontal cross-section B-B'

**RADIATION SHIELD DESCRIPTION**
**P1.** Exhaust fan with photovoltaic panel. Fixed on a white water-resistant ABS cover. Exhaust capacity: 20 m$^3$/hour.
**P2.** White water-resistant PVC body structure.
**P3.** Screwed connection between exhaust fan and body structure. 3x zinc-plated steel, round head screw with self-locking nut. Ø 4mm L=20 mm.
**P4.** Temperature and relative humidity sensor.
**P5.** Plastic clamp.
**P6.** Structure for suspending the sensor inside the radiation shield body. 2x zinc-plated steel, threaded rod with self-locking nuts and washers. Ø3 mm, L=150 mm.
**P7.** Interior black EVA foam coating to increase thermal insulation. 5 mm thick.

*Metadata associated to each measuring site:*

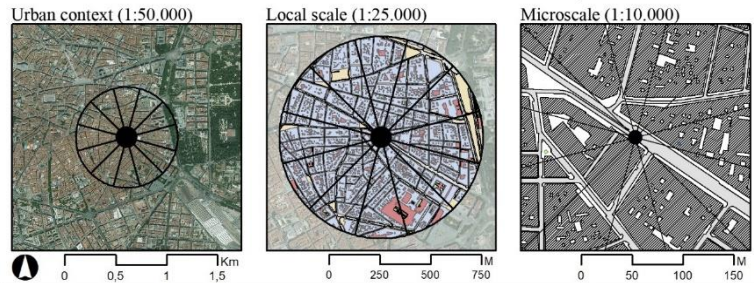

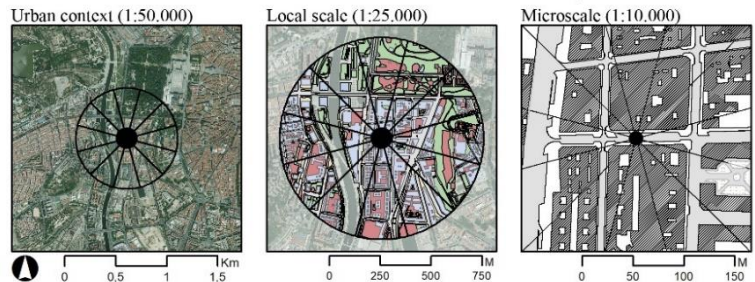

## Site 03 LA CHOPERA — Calle de Embajadores, 168. 28045 Madrid

Urban context (1:50.000)
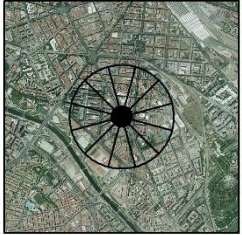
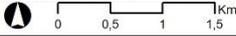

Local scale (1:25.000)
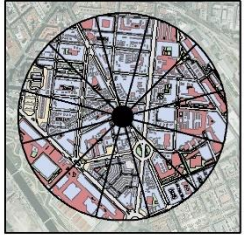
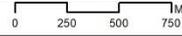

Microscale (1:10.000)
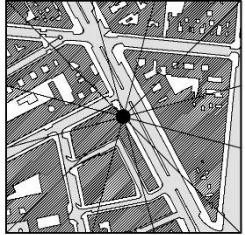
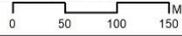

**SITE DESCRIPTION**
District: **02 - Arganzuela**
Neigbourh.: **023 - La Chopera**
Lat: **40.396**   Long: **-3.696**
LCZ: **02 - Compact midrise**
SVF summer/winter: **0.4/0.6**
Aspect ratio: **1.3**

**LAND COVER**
Building s.f.: **44%**
Impervious s.f.: **47%**
Pervious s.f.: **9%**
Typical buildings height: **20 m**
Typical tree height: **6-12 m**
Davenport roughness class: **7**
Traffic density: **6/10**
Heat pumps to street: **0.67**
Typical road materials: **Asphalt**
Typical wall materials: **Bricks**

**SENSOR DESCRIPTION**
Mast type: **Streetlight**
Sensor height: **6 m**
Radiation shield: **Yes**
Mechanical ventilation: **Yes**
Parameters: **D. b. temperature (ºC)**
**Relative humidity (%)**

Surface cover by sectors
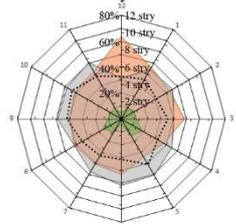

Sensor location
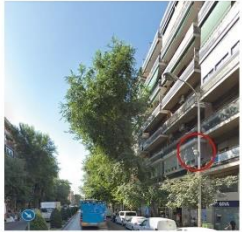

Sky View Factor in summer
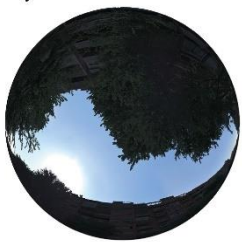

Sky View Factor in winter
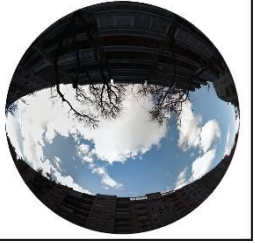

## Site 04 ESTRELLA — Calle de la Estrella Polar, 8. 28007 Madrid

Urban context (1:50.000)
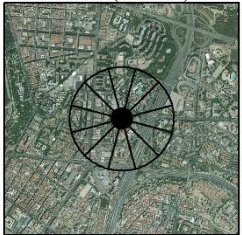
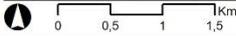

Local scale (1:25.000)
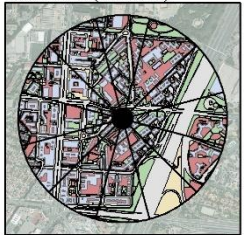
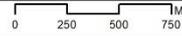

Microscale (1:10.000)
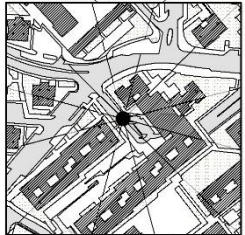
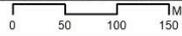

**SITE DESCRIPTION**
District: **03 - Retiro**
Neigbourh.: **033 - Estrella**
Lat: **40.411**   Long: **-3.667**
LCZ: **04 - Open high-rise**
SVF summer/winter: **0.6/0.7**
Aspect ratio: **1.4**

**LAND COVER**
Building s.f.: **20%**
Impervious s.f.: **52%**
Pervious s.f.: **28%**
Typical buildings height: **28-52 m**
Typical tree height: **12 m**
Davenport roughness class: **8**
Traffic density: **1/10**
Heat pumps to street: **0.52**
Typical road materials: **Asphalt**
Typical wall materials: **Bricks**

**SENSOR DESCRIPTION**
Mast type: **Streetlight**
Sensor height: **6 m**
Radiation shield: **Yes**
Mechanical ventilation: **Yes**
Parameters: **D. b. temperature (ºC)**
**Relative humidity (%)**

Surface cover by sectors
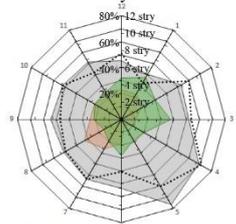

Sensor location
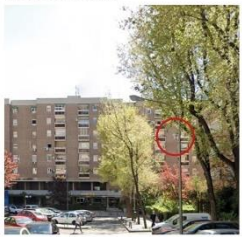

Sky View Factor in summer
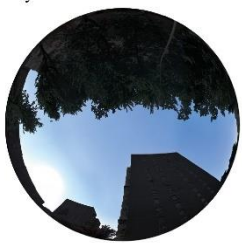

Sky View Factor in winter
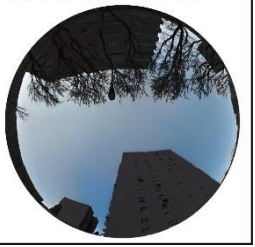

## Site 05 PACÍFICO — Calle de Granada, 12. 28007 Madrid

Urban context (1:50.000)
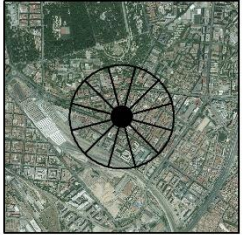
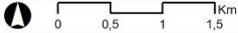

Local scale (1:25.000)
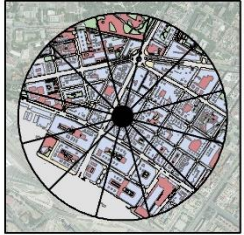
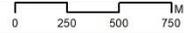

Microscale (1:10.000)
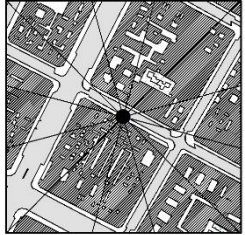
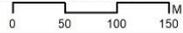

**SITE DESCRIPTION**
District: **03 - Retiro**
Neigbourh.: **031 - Pacífico**
Lat: **40.405**  Long: **-3.680**
LCZ: **02 - Compact midrise**
SVF summer/winter: **0.4/0.5**
Aspect ratio: **1.7**

**LAND COVER**
Building s.f.: **39%**
Impervious s.f.: **51%**
Pervious s.f.: **10%**
Typical buildings height: **26 m**
Typical tree height: **8 m**
Davenport roughness class: **7**
Traffic density: **1/10**
Heat pumps to street: **0.50**
Typical road materials: **Asphalt**
Typical wall materials: **Bricks**

**SENSOR DESCRIPTION**
Mast type: **Streetlight**
Sensor height: **6 m**
Radiation shield: **Yes**
Mechanical ventilation: **Yes**
Parameters: **D. b. temperature (°C)**
**Relative humidity (%)**

Surface cover by sectors
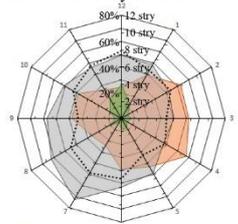

Sensor location
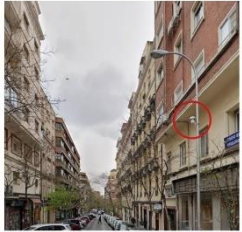

Sky View Factor in summer
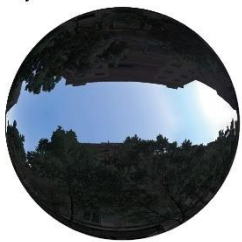

Sky View Factor in winter
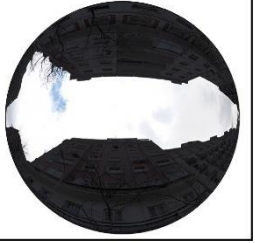

## Site 06 GUINDALERA — Calle de Eraso, 34. 28028 Madrid

Urban context (1:50.000)
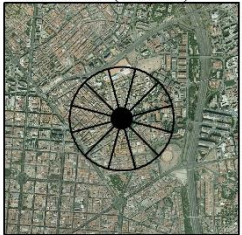
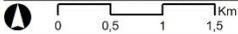

Local scale (1:25.000)
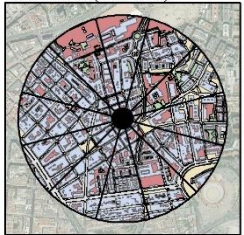
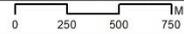

Microscale (1:10.000)
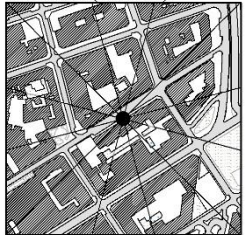
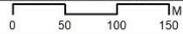

**SITE DESCRIPTION**
District: **04 - Salamanca**
Neigbourh.: **044 - Guindalera**
Lat: **40.435**  Long: **-3.668**
LCZ: **02 - Compact midrise**
SVF summer/winter: **0.6/0.6**
Aspect ratio: **1.5**

**LAND COVER**
Building s.f.: **41%**
Impervious s.f.: **47%**
Pervious s.f.: **12%**
Typical buildings height: **15 m**
Typical tree height: **-**
Davenport roughness class: **7**
Traffic density: **3/10**
Heat pumps to street: **0.22**
Typical road materials: **Asphalt**
Typical wall materials: **Bricks**

**SENSOR DESCRIPTION**
Mast type: **Streetlight**
Sensor height: **6 m**
Radiation shield: **Yes**
Mechanical ventilation: **Yes**
Parameters: **D. b. temperature (°C)**
**Relative humidity (%)**

Surface cover by sectors
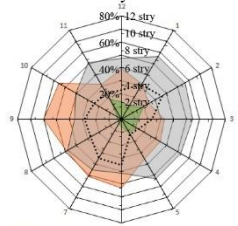

Sensor location
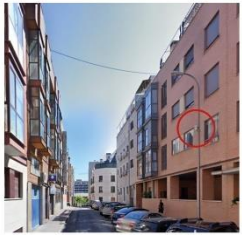

Sky View Factor in summer
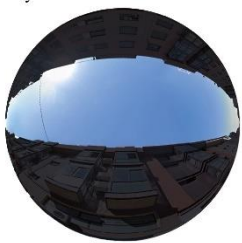

Sky View Factor in winter
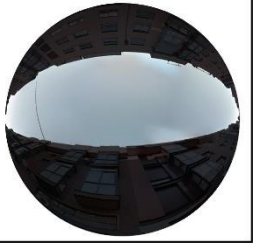

## Site 07 RECOLETOS — Calle de Goya, 14. 28001 Madrid

Urban context (1:50.000)
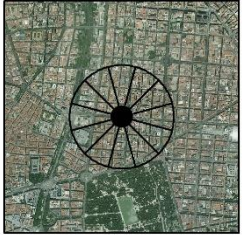
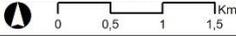

Local scale (1:25.000)
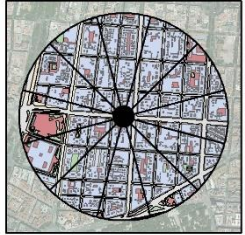
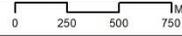

Microscale (1:10.000)
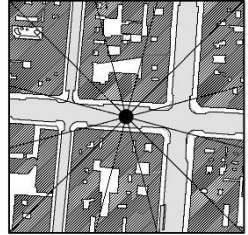
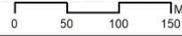

**SITE DESCRIPTION**
District: **04 - Salamanca**
Neigbourh.: **041 - Recoletos**
Lat: **40.425**   Long: **-3.685**
LCZ: **02 - Compact midrise**
SVF summer/winter: **0.7/0.8**
Aspect ratio: **1.3**

**LAND COVER**
Building s.f.: **52%**
Impervious s.f.: **42%**
Pervious s.f.: **6%**
Typical buildings height: **20 m**
Typical tree height: **8-12 m**
Davenport roughness class: **7**
Traffic density: **9/10**
Heat pumps to street: **0.00**
Typical road materials: **Asphalt**
Typical wall materials: **Bricks**

**SENSOR DESCRIPTION**
Mast type: **Streetlight**
Sensor height: **6 m**
Radiation shield: **Yes**
Mechanical ventilation: **Yes**
Parameters: **D. b. temperature (°C)**
            **Relative humidity (%)**

Surface cover by sectors
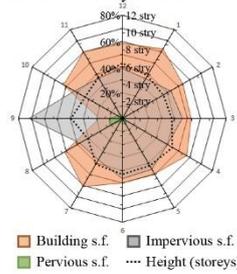

Sensor location
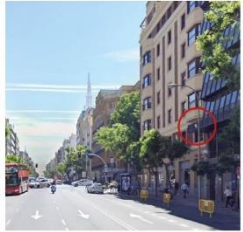

Sky View Factor in summer
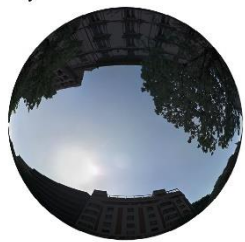

Sky View Factor in winter
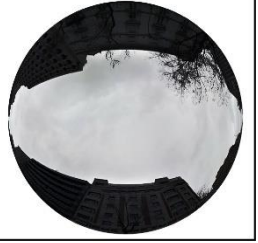

## Site 08 HISPANOAMÉRICA — Calle Infanta María Teresa, 12. 28016 Madrid

Urban context (1:50.000)
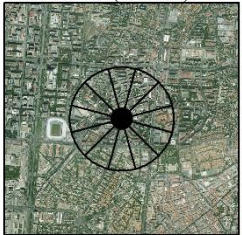
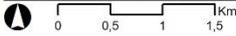

Local scale (1:25.000)
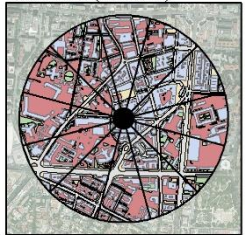
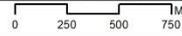

Microscale (1:10.000)
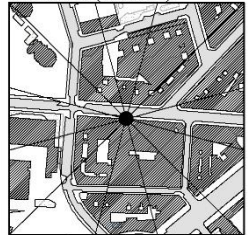
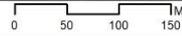

**SITE DESCRIPTION**
District: **05 - Chamartín**
Neigbourh.: **054 - Hispanoamérica**
Lat: **40.454**   Long: **-3.681**
LCZ: **02 - Compact midrise**
SVF summer/winter: **0.6/0.6**
Aspect ratio: **1.1**

**LAND COVER**
Building s.f.: **29%**
Impervious s.f.: **49%**
Pervious s.f.: **22%**
Typical buildings height: **16 m**
Typical tree height: **12 m**
Davenport roughness class: **7**
Traffic density: **6/10**
Heat pumps to street: **0.18**
Typical road materials: **Asphalt**
Typical wall materials: **Bricks**

**SENSOR DESCRIPTION**
Mast type: **Streetlight**
Sensor height: **6 m**
Radiation shield: **Yes**
Mechanical ventilation: **Yes**
Parameters: **D. b. temperature (°C)**
            **Relative humidity (%)**

Surface cover by sectors
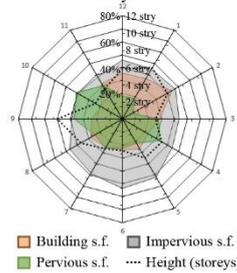

Sensor location
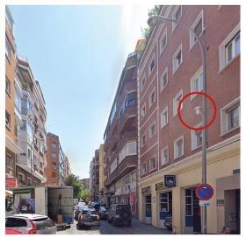

Sky View Factor in summer
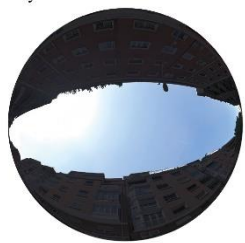

Sky View Factor in winter
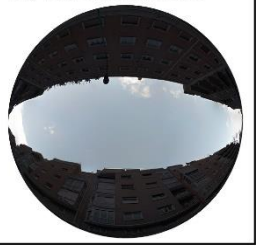

## Site 09 CUATRO CAMINOS — Calle de Juan de Olías, 4. 28020 Madrid

Urban context (1:50.000)
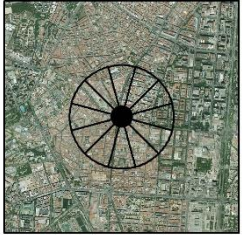
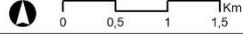

Local scale (1:25.000)
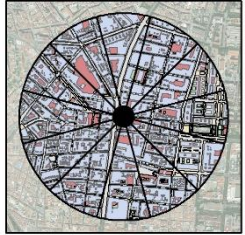
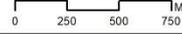

Microscale (1:10.000)
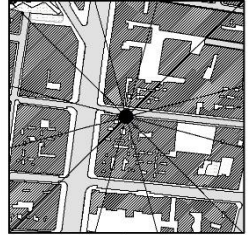
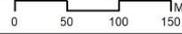

**SITE DESCRIPTION**
District: 06 - Tetuán
Neigbourh.: 062 - Cuatro Caminos
Lat: 40.453  Long: -3.703
LCZ: 02 - Compact midrise
SVF summer/winter: 0.6/0.6
Aspect ratio: 1.2

**LAND COVER**
Building s.f.: 53%
Impervious s.f.: 41%
Pervious s.f.: 6%
Typical buildings height: 12 m
Typical tree height: -
Davenport roughness class: 7
Traffic density: 2/10
Heat pumps to street: 0.23
Typical road materials: Asphalt
Typical wall materials: Bricks

**SENSOR DESCRIPTION**
Mast type: Streetlight
Sensor height: 6 m
Radiation shield: Yes
Mechanical ventilation: Yes
Parameters: D. b. temperature (ºC)
Relative humidity (%)

Surface cover by sectors
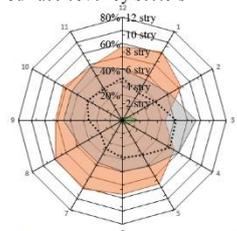

Sensor location
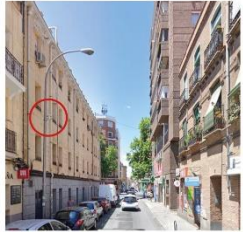

Sky View Factor in summer
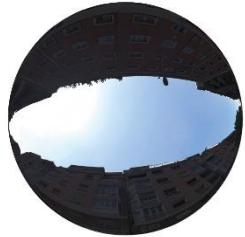

Sky View Factor in winter
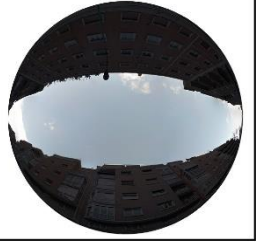

## Site 10 ARAPILES — Calle de Fernández de los Ríos, 25. 28015 Madrid

Urban context (1:50.000)
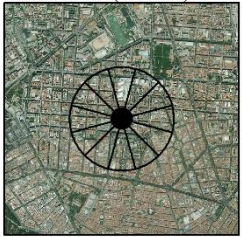
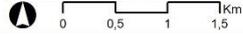

Local scale (1:25.000)
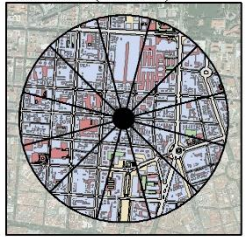
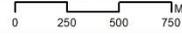

Microscale (1:10.000)
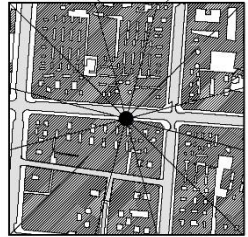
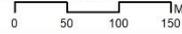

**SITE DESCRIPTION**
District: 07 - Chamberí
Neigbourh.: 072 - Arapiles
Lat: 40.435  Long: -3.707
LCZ: 02 - Compact midrise
SVF summer/winter: 0.5/0.6
Aspect ratio: 1.3

**LAND COVER**
Building s.f.: 53%
Impervious s.f.: 42%
Pervious s.f.: 5%
Typical buildings height: 20 m
Typical tree height: 5-12 m
Davenport roughness class: 7
Traffic density: 2/10
Heat pumps to street: 0.20
Typical road materials: Asphalt
Typical wall materials: Bricks

**SENSOR DESCRIPTION**
Mast type: Streetlight
Sensor height: 6 m
Radiation shield: Yes
Mechanical ventilation: Yes
Parameters: D. b. temperature (ºC)
Relative humidity (%)

Surface cover by sectors
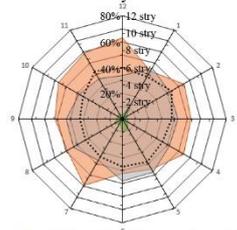

Sensor location
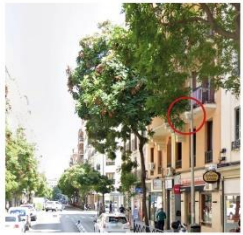

Sky View Factor in summer
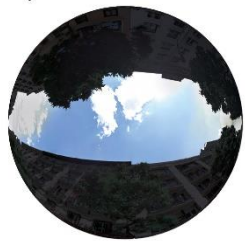

Sky View Factor in winter
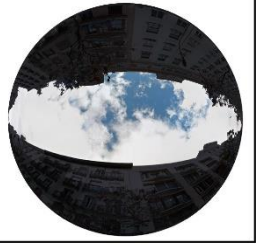

## Site 11 PEÑAGRANDE — Calle de las Islas Cíes, 5. 28035 Madrid

Urban context (1:50.000)
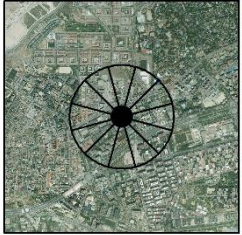
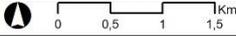

Local scale (1:25.000)
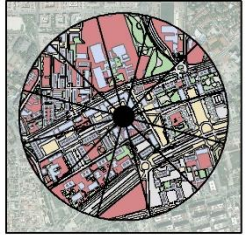
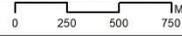

Microscale (1:10.000)
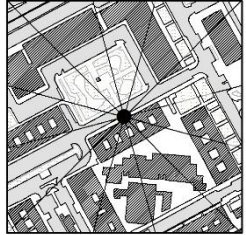
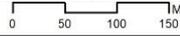

**SITE DESCRIPTION**
District: **08 - Fuencarral-El Pardo**
Neigbourh.: **083 - Peñagrande**
Lat: **40.481**  Long: **-3.718**
LCZ: **04 - Open high-rise**
SVF summer/winter: **0.7/0.7**
Aspect ratio: **0.8**

**LAND COVER**
Building s.f.: **19%**
Impervious s.f.: **54%**
Pervious s.f.: **28%**
Typical buildings height: **30 m**
Typical tree height: **5 m**
Davenport roughness class: **7**
Traffic density: **3/10**
Heat pumps to street: **0.49**
Typical road materials: **Asphalt**
Typical wall materials: **Bricks**

**SENSOR DESCRIPTION**
Mast type: **Streetlight**
Sensor height: **6 m**
Radiation shield: **Yes**
Mechanical ventilation: **Yes**
Parameters: **D. b. temperature (°C)**
**Relative humidity (%)**

Surface cover by sectors
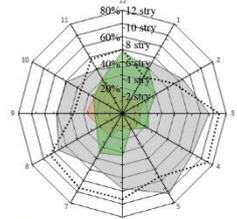

Sensor location
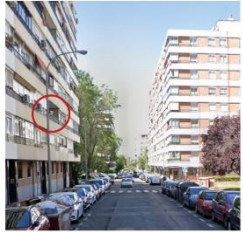

Sky View Factor in summer
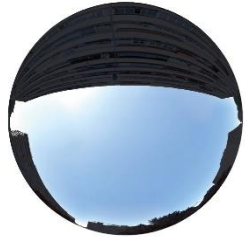

Sky View Factor in winter
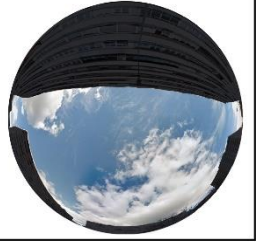

## Site 12 ARAVACA — Avenida de la Osa Mayor, 15. 28023 Madrid

Urban context (1:50.000)
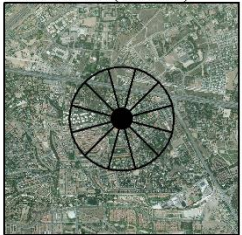
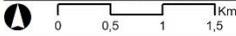

Local scale (1:25.000)
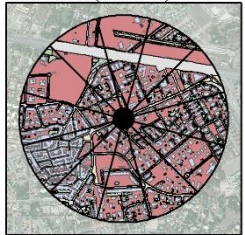
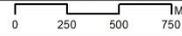

Microscale (1:10.000)
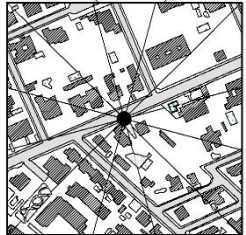
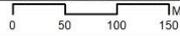

**SITE DESCRIPTION**
District: **09 - Moncloa-Aravaca**
Neigbourh.: **097 - Aravaca**
Lat: **40.460**  Long: **-3.777**
LCZ: **06 - Open low-rise**
SVF summer/winter: **0.9/1.0**
Aspect ratio: **0.4**

**LAND COVER**
Building s.f.: **20%**
Impervious s.f.: **37%**
Pervious s.f.: **43%**
Typical buildings height: **8 m**
Typical tree height: **4-12 m**
Davenport roughness class: **5**
Traffic density: **1/10**
Heat pumps to street: **0.00**
Typical road materials: **Asphalt**
Typical wall materials: **Bricks**

**SENSOR DESCRIPTION**
Mast type: **Streetlight**
Sensor height: **6 m**
Radiation shield: **Yes**
Mechanical ventilation: **Yes**
Parameters: **D. b. temperature (°C)**
**Relative humidity (%)**

Surface cover by sectors
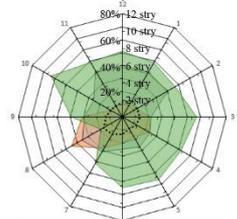

Sensor location
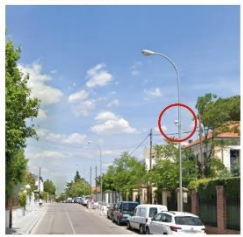

Sky View Factor in summer
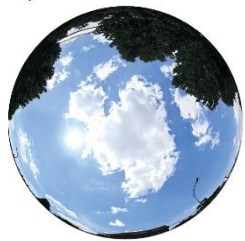

Sky View Factor in winter
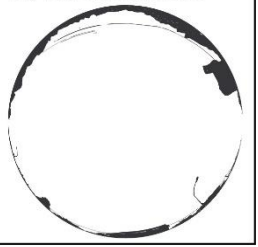

## Site 13 LOS CÁRMENES — Calle de Gallur, 320. 28047 Madrid

Urban context (1:50.000)
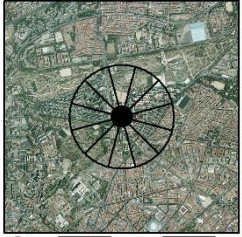
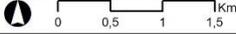

Local scale (1:25.000)
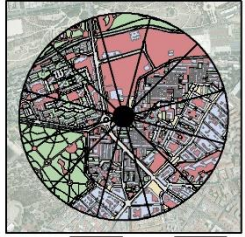
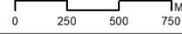

Microscale (1:10.000)
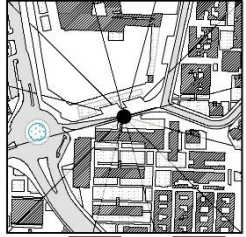
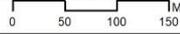

**SITE DESCRIPTION**
District: **10 - La Latina**
Neigbourh.: **101 - Los Cármenes**
Lat: **40.396**  Long: **-3.743**
LCZ: **05 - Open midrise**
SVF summer/winter: **0.9/1.0**
Aspect ratio: **0.8**

**LAND COVER**
Building s.f.: **21%**
Impervious s.f.: **33%**
Pervious s.f.: **45%**
Typical buildings height: **22 m**
Typical tree height: **12 m**
Davenport roughness class: **6**
Traffic density: **5/10**
Heat pumps to street: **0.13**
Typical road materials: **Asphalt**
Typical wall materials: **Bricks**

**SENSOR DESCRIPTION**
Mast type: **Streetlight**
Sensor height: **6 m**
Radiation shield: **Yes**
Mechanical ventilation: **Yes**
Parameters: **D. b. temperature (°C)**
**Relative humidity (%)**

Surface cover by sectors
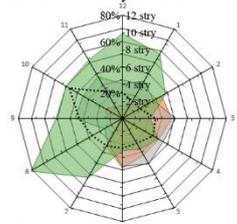

Sensor location
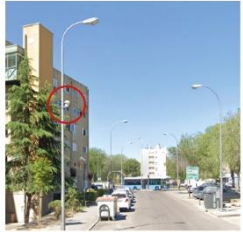

Sky View Factor in summer
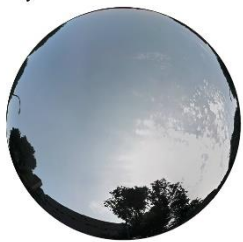

Sky View Factor in winter
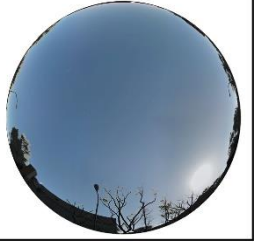

## Site 14 OPAÑEL — Calle del Arroyo Opañel, 19. 28019 Madrid

Urban context (1:50.000)
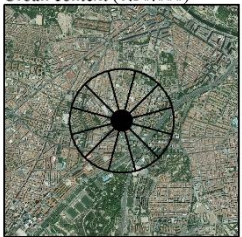
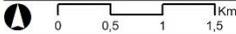

Local scale (1:25.000)
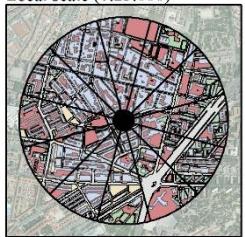
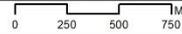

Microscale (1:10.000)
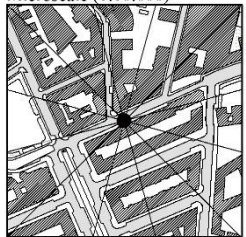
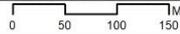

**SITE DESCRIPTION**
District: **11 - Carabanchel**
Neigbourh.: **112 - Opañel**
Lat: **40.388**  Long: **-3.719**
LCZ: **05 - Open midrise**
SVF summer/winter: **0.8/0.8**
Aspect ratio: **1.1**

**LAND COVER**
Building s.f.: **31%**
Impervious s.f.: **48%**
Pervious s.f.: **21%**
Typical buildings height: **14 m**
Typical tree height: **-**
Davenport roughness class: **7**
Traffic density: **2/10**
Heat pumps to street: **0.64**
Typical road materials: **Asphalt**
Typical wall materials: **Bricks**

**SENSOR DESCRIPTION**
Mast type: **Streetlight**
Sensor height: **6 m**
Radiation shield: **Yes**
Mechanical ventilation: **Yes**
Parameters: **D. b. temperature (°C)**
**Relative humidity (%)**

Surface cover by sectors
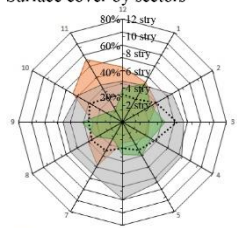

Sensor location
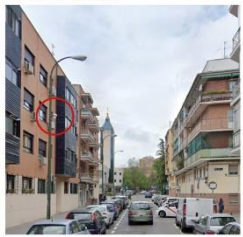

Sky View Factor in summer
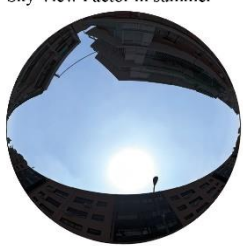

Sky View Factor in winter
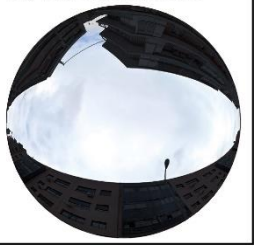

## Site 15 SAN DIEGO — Calle del Monte Perdido, 82. 28053 Madrid

Urban context (1:50.000)
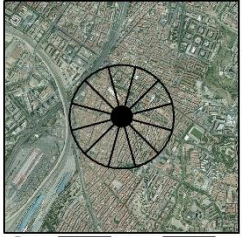
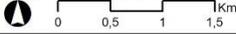

Local scale (1:25.000)
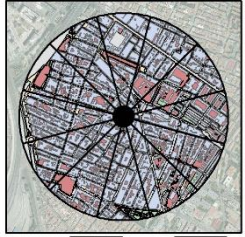
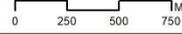

Microscale (1:10.000)
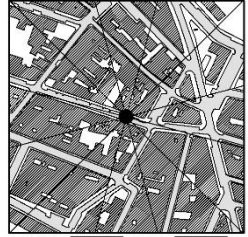
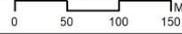

**SITE DESCRIPTION**
District: **13 - Puente de Vallecas**
Neigbourh.: **132 - San Diego**
Lat: **40.393**   Long: **-3.667**
LCZ: **02 - Compact midrise**
SVF summer/winter: **0.6/0.6**
Aspect ratio: **1.1**

**LAND COVER**
Building s.f.: **48%**
Impervious s.f.: **45%**
Pervious s.f.: **7%**
Typical buildings height: **11 m**
Typical tree height: **5 m**
Davenport roughness class: **7**
Traffic density: **2/10**
Heat pumps to street: **0.30**
Typical road materials: **Asphalt**
Typical wall materials: **Bricks**

**SENSOR DESCRIPTION**
Mast type: **Streetlight**
Sensor height: **6 m**
Radiation shield: **Yes**
Mechanical ventilation: **Yes**
Parameters: **D. b. temperature (°C)**
   **Relative humidity (%)**

Surface cover by sectors
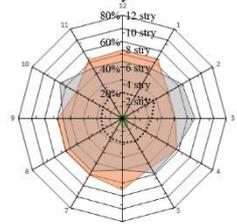

Sensor location
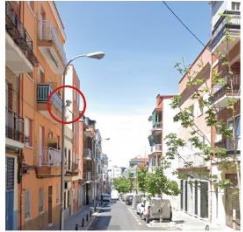

Sky View Factor in summer
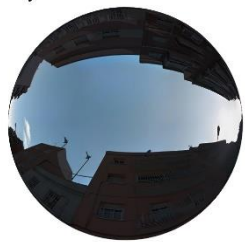

Sky View Factor in winter
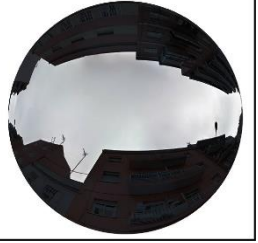

---

## Site 16 HORCAJO — Calle de la Provenza, 13F. 28030 Madrid

Urban context (1:50.000)
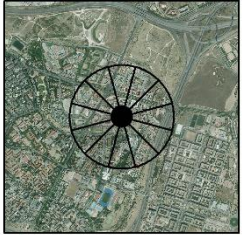
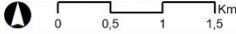

Local scale (1:25.000)
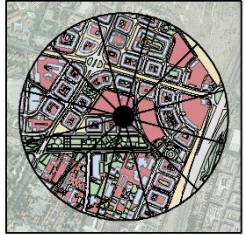
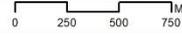

Microscale (1:10.000)
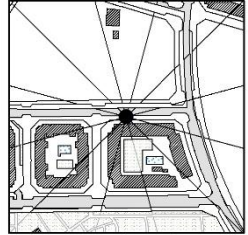
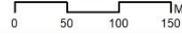

**SITE DESCRIPTION**
District: **14 - Moratalaz**
Neigbourh.: **142 - Horcajo**
Lat: **40.405**   Long: **-3.630**
LCZ: **09 - Sparsely built**
SVF summer/winter: **0.9/0.9**
Aspect ratio: **0.2**

**LAND COVER**
Building s.f.: **17%**
Impervious s.f.: **49%**
Pervious s.f.: **35%**
Typical buildings height: **20 m**
Typical tree height: **7 m**
Davenport roughness class: **6**
Traffic density: **1/10**
Heat pumps to street: **0.00**
Typical road materials: **Asphalt**
Typical wall materials: **Bricks**

**SENSOR DESCRIPTION**
Mast type: **Streetlight**
Sensor height: **6 m**
Radiation shield: **Yes**
Mechanical ventilation: **Yes**
Parameters: **D. b. temperature (°C)**
   **Relative humidity (%)**

Surface cover by sectors
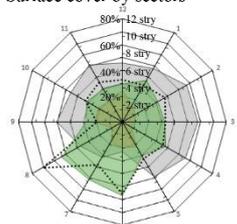

Sensor location
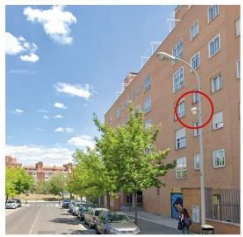

Sky View Factor in summer
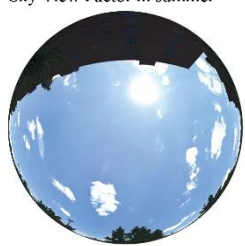

Sky View Factor in winter
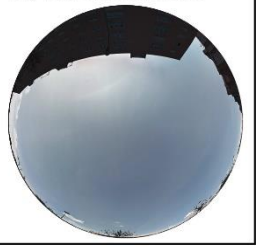

## Site 17 PUEBLO NUEVO — Calle de Vital Aza, 30. 28017 Madrid

Urban context (1:50.000)
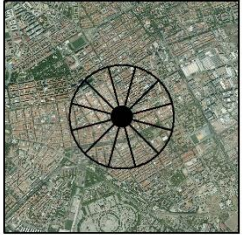
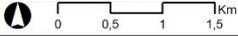

Local scale (1:25.000)
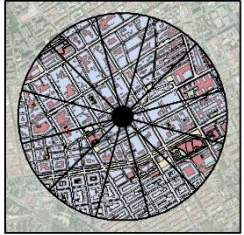
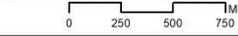

Microscale (1:10.000)
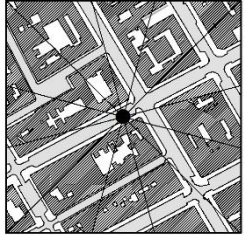
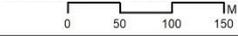

**SITE DESCRIPTION**
District: **15 - Ciudad Lineal**
Neigbourh.: **152 - Pueblo Nuevo**
Lat: **40.432**   Long: **-3.643**
LCZ: **02 - Compact midrise**
SVF summer/winter: **0.8/0.8**
Aspect ratio: **0.9**

**LAND COVER**
Building s.f.: **44%**
Impervious s.f.: **50%**
Pervious s.f.: **6%**
Typical buildings height: **17 m**
Typical tree height: **5-12 m**
Davenport roughness class: **7**
Traffic density: **2/10**
Heat pumps to street: **0.50**
Typical road materials: **Asphalt**
Typical wall materials: **Bricks**

**SENSOR DESCRIPTION**
Mast type: **Streetlight**
Sensor height: **6 m**
Radiation shield: **Yes**
Mechanical ventilation: **Yes**
Parameters: **D. b. temperature (ºC)**
   **Relative humidity (%)**

Surface cover by sectors
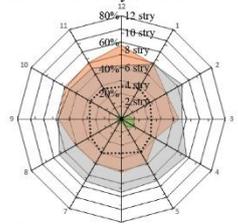
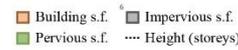

Sensor location
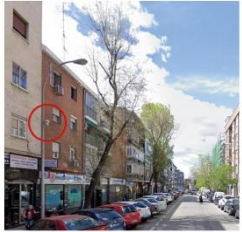

Sky View Factor in summer
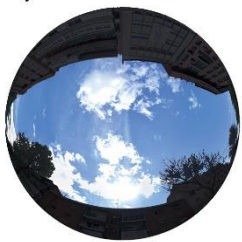

Sky View Factor in winter
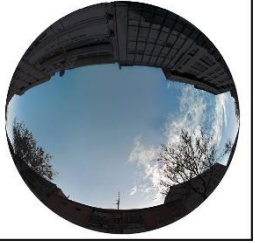

---

## Site 18 CANILLAS — Calle de Santa Natalia, 5. 28043 Madrid

Urban context (1:50.000)
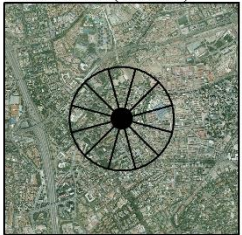
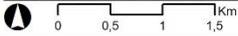

Local scale (1:25.000)
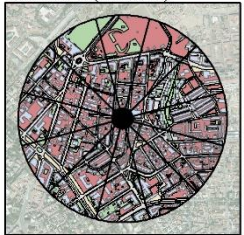
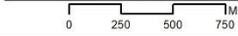

Microscale (1:10.000)
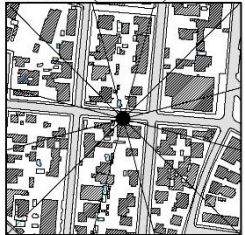
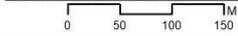

**SITE DESCRIPTION**
District: **16 - Hortaleza**
Neigbourh.: **163 - Canillas**
Lat: **40.461**   Long: **-3.655**
LCZ: **06 - Open low-rise**
SVF summer/winter: **0.4/0.8**
Aspect ratio: **0.9**

**LAND COVER**
Building s.f.: **22%**
Impervious s.f.: **46%**
Pervious s.f.: **32%**
Typical buildings height: **10 m**
Typical tree height: **8-20 m**
Davenport roughness class: **7**
Traffic density: **1/10**
Heat pumps to street: **0.00**
Typical road materials: **Asphalt**
Typical wall materials: **Bricks**

**SENSOR DESCRIPTION**
Mast type: **Streetlight**
Sensor height: **6 m**
Radiation shield: **Yes**
Mechanical ventilation: **Yes**
Parameters: **D. b. temperature (ºC)**
   **Relative humidity (%)**

Surface cover by sectors
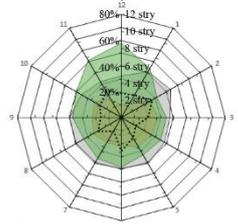
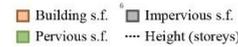

Sensor location
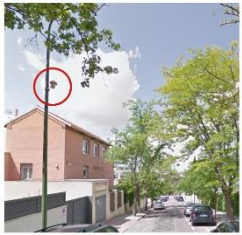

Sky View Factor in summer
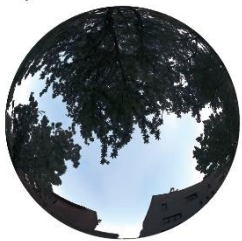

Sky View Factor in winter
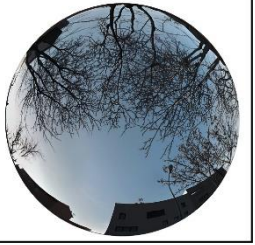

## Site 19 LOS ÁNGELES — Calle la del Soto del Parral, 15. 28041 Madrid

Urban context (1:50.000)
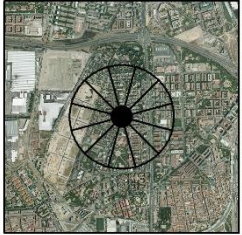
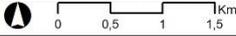

Local scale (1:25.000)
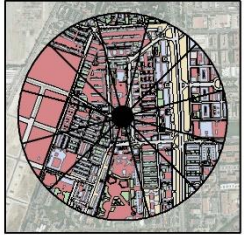
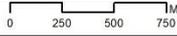

Microscale (1:10.000)
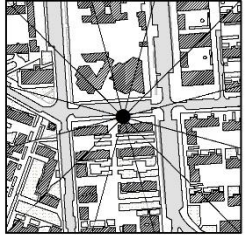
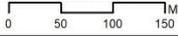

**SITE DESCRIPTION**
District: **17 - Villaverde**
Neigbourh.: **175 - Los Ángeles**
Lat: **40.358**   Long: **-3.696**
LCZ: **05 - Open midrise**
SVF summer/winter: **0.9/0.9**
Aspect ratio: **0.8**

**LAND COVER**
Building s.f.: **17%**
Impervious s.f.: **49%**
Pervious s.f.: **34%**
Typical buildings height: **17 m**
Typical tree height: **8 m**
Davenport roughness class: **6**
Traffic density: **2/10**
Heat pumps to street: **0.00**
Typical road materials: **Asphalt**
Typical wall materials: **Bricks**

**SENSOR DESCRIPTION**
Mast type: **Streetlight**
Sensor height: **6 m**
Radiation shield: **Yes**
Mechanical ventilation: **Yes**
Parameters: **D. b. temperature (ºC)**
            **Relative humidity (%)**

Surface cover by sectors
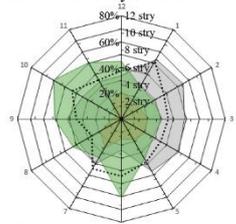

Sensor location
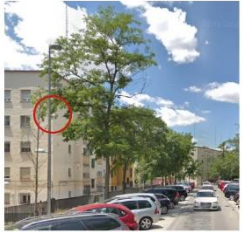

Sky View Factor in summer
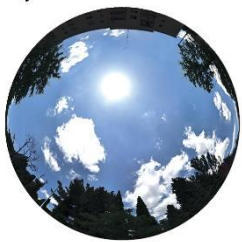

Sky View Factor in winter
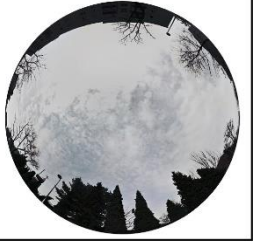

## Site 20 CANILLEJAS — Calle de San Narciso, 19. 28022 Madrid

Urban context (1:50.000)
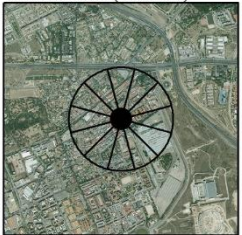
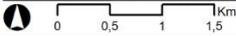

Local scale (1:25.000)
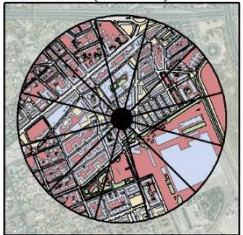
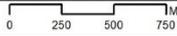

Microscale (1:10.000)
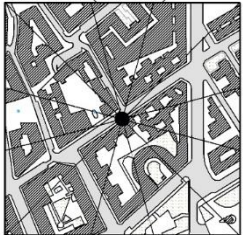
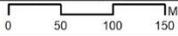

**SITE DESCRIPTION**
District: **20 - San Blas-Canillejas**
Neigbourh.: **207 - Canillejas**
Lat: **40.445**   Long: **-3.610**
LCZ: **05 - Open midrise**
SVF summer/winter: **0.8/0.8**
Aspect ratio: **1.2**

**LAND COVER**
Building s.f.: **31%**
Impervious s.f.: **49%**
Pervious s.f.: **20%**
Typical buildings height: **14 m**
Typical tree height: **-**
Davenport roughness class: **7**
Traffic density: **2/10**
Heat pumps to street: **0.34**
Typical road materials: **Asphalt**
Typical wall materials: **Bricks**

**SENSOR DESCRIPTION**
Mast type: **Streetlight**
Sensor height: **6 m**
Radiation shield: **Yes**
Mechanical ventilation: **Yes**
Parameters: **D. b. temperature (ºC)**
            **Relative humidity (%)**

Surface cover by sectors
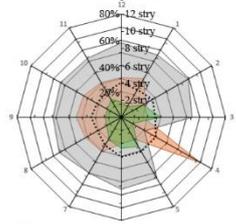

Sensor location
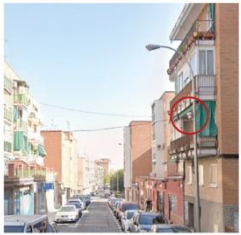

Sky View Factor in summer
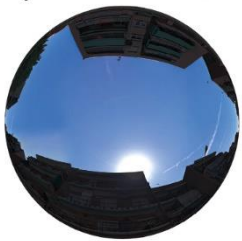

Sky View Factor in winter
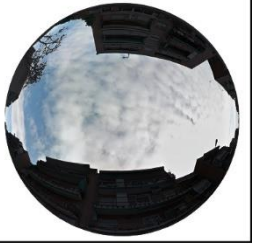

**Funding**

This research was funded by the FPU research grant FPU15/05052, from the Spanish Ministry of Science, Innovation and Universities. This research was also supported by the MODIFICA research project (BIA2013-41732-R), funded by the Spanish Ministry of Economy and Competitiveness.

**Acknowledgements**

The authors would like to thank the Spanish State Meteorological Agency (AEMET) for providing access to their meteorological data, and the Madrid City Council' Subdivision of Energy and Climate Change for their support with the urban measurements campaign. The authors would also like to thank Inês Costa Carrapiço for her valuable recommendations and inputs while proof-reading this manuscript.

**References**

AEMET, 2019. Informe anual 2018. Madrid, Spain.

Alexander, P.J., Bechtel, B., Chow, W.T.L., Fealy, R., Mills, G., 2016. Linking urban climate classification with an urban energy and water budget model: Multi-site and multi-seasonal evaluation. Urban Clim. 17, 196–215. https://doi.org/10.1016/j.uclim.2016.08.003

Alexander, P.J., Mills, G., 2014. Local climate classification and Dublin's urban heat island. Atmosphere (Basel). 5, 755–774. https://doi.org/10.3390/atmos5040755

Ampatzidis, P., Kershaw, T., 2020. A review of the impact of blue space on the urban microclimate. Sci. Total Environ. 730, 139068. https://doi.org/10.1016/j.scitotenv.2020.139068

Ando, T., Ueyama, M., 2017. Surface energy exchange in a dense urban built-up area based on two-year eddy covariance measurements in Sakai, Japan. Urban Clim. 19, 155–169. https://doi.org/10.1016/j.uclim.2017.01.005

Arnds, D., Boehner, J., Bechtel, B., 2017. Spatio-temporal variance and meteorological drivers of the urban heat island in a European city. Theor. Appl. Climatol. 128, 43–61. https://doi.org/10.1007/s00704-015-1687-4

Ayuntamiento de Madrid, 2015. Cartografía municipal por distritos a escala 1:1000. Portal de datos abiertos.

Ayuntamiento de Madrid, 2013. Tráfico. Intensidad media diaria anual por tramos. Portal de datos abiertos.

Barlow, J.F., 2014. Progress in observing and modelling the urban boundary layer. Urban Clim. 10, 216–240. https://doi.org/10.1016/j.uclim.2014.03.011

Bechtel, B., Alexander, P.J., Beck, C., Böhner, J., Brousse, O., Ching, J., Demuzere, M., Fonte, C., Gál, T., Hidalgo, J., Hoffmann, P., Middel, A., Mills, G., Ren, C., See, L., Sismanidis, P., Verdonck, M.L., Xu, G., Xu, Y., 2019. Generating WUDAPT Level 0 data – Current status of production and evaluation. Urban Clim. 27, 24–45. https://doi.org/10.1016/j.uclim.2018.10.001

Bechtel, B., Alexander, P.J., Böhner, J., Ching, J., Conrad, O., Feddema, J., Mills, G., See, L., Stewart, I.D., 2015. Mapping Local Climate Zones for a Worldwide Database of the Form and Function of Cities. ISPRS Int. J. Geo-Information 4, 199–219. https://doi.org/10.3390/ijgi4010199

Beck, C., Straub, A., Breitner, S., Cyrys, J., Philipp, A., Rathmann, J., Schneider, A., Wolf, K., Jacobeit,

J., 2018. Air temperature characteristics of local climate zones in the Augsburg urban area (Bavaria, southern Germany) under varying synoptic conditions. Urban Clim. 25, 152–166. https://doi.org/10.1016/j.uclim.2018.04.007

Bell, S., Cornford, D., Bastin, L., 2015. How good are citizen weather stations? Addressing a biased opinion. Weather 70, 75–84. https://doi.org/10.1002/wea.2316

Best, M.J., Grimmmmond, C.S.B., 2015. Key conclusions of the first international urban land surface model comparison project. Bull. Am. Meteorol. Soc. 96, 805–819. https://doi.org/10.1175/BAMS-D-14-00122.1

Borbora, J., Das, A.K., 2014. Summertime Urban Heat Island study for Guwahati City, India. Sustain. Cities Soc. 11, 61–66. https://doi.org/10.1016/j.scs.2013.12.001

Brousse, O., Martilli, A., Foley, M., Mills, G., Bechtel, B., 2016. WUDAPT, an efficient land use producing data tool for mesoscale models? Integration of urban LCZ in WRF over Madrid. Urban Clim. 17, 116–134. https://doi.org/10.1016/j.uclim.2016.04.001

Budhiraja, B., Agrawal, G., Pathak, P., 2020. Urban heat island effect of a polynuclear megacity Delhi – Compactness and thermal evaluation of four sub-cities. Urban Clim. 32, 100634. https://doi.org/10.1016/j.uclim.2020.100634

Chandler, T.J., 1965. The climate of London. Hutchinson of London, London.

Chapman, L., Bell, C., Bell, S., 2017. Can the crowdsourcing data paradigm take atmospheric science to a new level? A case study of the urban heat island of London quantified using Netatmo weather stations. Int. J. Climatol. 37, 3597–3605. https://doi.org/10.1002/joc.4940

Ching, J.K.S., 2013. A perspective on urban canopy layer modeling for weather, climate and air quality applications. Urban Clim. 3, 13–39. https://doi.org/10.1016/j.uclim.2013.02.001

Christen, A., 2014. Atmospheric measurement techniques to quantify greenhouse gas emissions from cities. Urban Clim. 10, 241–260. https://doi.org/10.1016/j.uclim.2014.04.006

Coseo, P., Larsen, L., 2014. How factors of land use/land cover, building configuration, and adjacent heat sources and sinks explain Urban Heat Islands in Chicago. Landsc. Urban Plan. 125, 117–129. https://doi.org/10.1016/j.landurbplan.2014.02.019

Crawford, B., Christen, A., 2015. Spatial source attribution of measured urban eddy covariance $CO_2$ fluxes. Theor. Appl. Climatol. 119, 733–755. https://doi.org/10.1007/s00704-014-1124-0

Doan, V.Q., Kusaka, H., Nguyen, T.M., 2019. Roles of past, present, and future land use and anthropogenic heat release changes on urban heat island effects in Hanoi, Vietnam: Numerical experiments with a regional climate model. Sustain. Cities Soc. 47, 101479. https://doi.org/10.1016/j.scs.2019.101479

Dodman, D., 2009. Blaming cities for climate change? An analysis of urban greenhouse gas emissions inventories. Environ. Urban. 21, 185–201. https://doi.org/10.1177/0956247809103016

Emmanuel, R., Krüger, E., 2012. Urban heat island and its impact on climate change resilience in a shrinking city: The case of Glasgow, UK. Build. Environ. 53, 137–149. https://doi.org/10.1016/j.buildenv.2012.01.020

Erell, E., Leal, V., Maldonado, E., 2005. Measurement of air temperature in the presence of a large radiant flux: An assessment of passively ventilated thermometer screens. Boundary-Layer Meteorol. 114, 205–231. https://doi.org/10.1007/s10546-004-8946-8

Erell, E., Williamson, T., 2007. Intra-urban differences in canopy layer air temperature at a mid-latitude city. Int. J. Climatol. 27, 1243–1255. https://doi.org/10.1002/joc.1469

Eurostat, 2020. Population on 1 January by age groups and sex - functional urban areas (URB_LPOP1).

Feigenwinter, C., Vogt, R., Christen, A., 2012. Eddy Covariance Measurements Over Urban Areas, in: Aubinet, M., Vesala, T., Papale, D. (Eds.), Eddy Covariance: A Practical Guide to Measurement and Data. Springer Science and Business Media, pp. 377–397. https://doi.org/10.1007/978-94-007-2351-1

Fenner, D., Meier, F., Bechtel, B., Otto, M., Scherer, D., 2017. Intra and inter "local climate zone" variability of air temperature as observed by crowdsourced citizen weather stations in Berlin, Germany. Meteorol. Zeitschrift 26, 525–547. https://doi.org/10.1127/metz/2017/0861

Fenner, D., Meier, F., Scherer, D., Polze, A., 2014. Spatial and temporal air temperature variability in Berlin, Germany, during the years 2001-2010. Urban Clim. 10, 308–331. https://doi.org/10.1016/j.uclim.2014.02.004

Fernández García, F., Allende Álvarez, F., Rasilla Álvarez, D., Martilli, A., Alcaide Muñoz, J., 2016. Estudio de detalle del clima urbano de Madrid. Madrid.

Fernández García, F., Almendros Coca, M.Á., López Gómez, A., 1996. La influencia del relieve en la isla de calor de Madrid: las vaguadas del Manzanares y del Abroñigal. Estud. Geográficos 57, 473–494.

Figuerola, P.I., Mazzeo, N. a., 1998. Urban-rural temperature differences in Buenos Aires. Int. J. Climatol. 18, 1709–1723. https://doi.org/10.1002/(SICI)1097-0088(199812)18:15<1709::AID-JOC338>3.0.CO;2-I

Gardes, T., Schoetter, R., Hidalgo, J., Long, N., Marquès, E., Masson, V., 2020. Statistical prediction of the nocturnal urban heat island intensity based on urban morphology and geographical factors - An investigation based on numerical model results for a large ensemble of French cities. Sci. Total Environ. 737, 139253. https://doi.org/10.1016/j.scitotenv.2020.139253

Geiger, R., 1950. The climate near the ground. Blue Hill Meteorological Observatory, Harvard University, Cambridge.

Grimmond, C.S.D., Blackett, M., Best, M.J., Baik, J.-J., Belcher, S.E., Beringer, J., Bohnenstengel, S.I., Calmet, I., Chen, F., Coutts, A., Dandou, A., Fortuniak, K., Gouvea, M.L., Hamdi, R., Hendry, M., Kanda, M., Kawai, T., Kawamoto, Y., Kondo, H., Krayenhoff, E.S., Lee, S.-H., Loridan, T., Martilli, A., Masson, V., Miao, S., Oleson, K., Ooka, R., Pigeon, G., Porson, A., Ryu, Y.-H., Salamanca, F., Steeneveld, G.-J., Tombrou, M., Voogt, J.A., Young, D.T., Zhang, N., 2011. Initial results from Phase 2 of the international urban energy balance model comparison. Int. J. Climatol. 31, 244–272. https://doi.org/10.1002/joc.2227

Grimmond, C.S.D., Blackett, M., Best, M.J., Barlow, J., Baik, J.-J., Belcher, S.E., Bohnenstengel, S.I., Calmet, I., Chen, F., Dandou, A., Fortuniak, K., Gouvea, M.L., Hamdi, R., Hendry, M., Kawai, T., Kawamoto, Y., Kondo, H., Krayenhoff, E.S., Lee, S.-H., Loridan, T., Martilli, A., Masson, V., Miao, S., Oleson, K., Pigeon, G., Porson, A., Ryu, Y.-H., Salamanca, F., Shashua-Bar, L., Steeneveld, G.-J., Tombrou, M., Voogt, J., Young, D., Zhang, N., 2010. The International Urban Energy Balance Models Comparison Project: First Results from Phase 1. J. Appl. Meteorol. Climatol. 49, 1268–1292. https://doi.org/10.1175/2010JAMC2354.1

Hammerberg, K., Brousse, O., Martilli, A., Mahdavi, A., 2018. Implications of employing detailed urban canopy parameters for mesoscale climate modelling: a comparison between WUDAPT and GIS databases over Vienna, Austria. Int. J. Climatol. 38, e1241–e1257. https://doi.org/10.1002/joc.5447

Hebbert, M., 2014. Climatology for city planning in historical perspective. Urban Clim. 10, 204–215. https://doi.org/10.1016/j.uclim.2014.07.001

Holmer, B., Thorsson, S., Eliasson, I., 2007. Cooling rates, sky view factors and the development of intra-urban air temperature difference. Geogr. Ann. Ser. A Phys. Geogr. 89 A, 237–248. https://doi.org/10.1111/j.1468-0459.2007.00323.x

Howard, L., 1833. The Climate of London. Harvey and Darton, London.

Jandaghian, Z., Berardi, U., 2020. Comparing urban canopy models for microclimate simulations in Weather Research and Forecasting Models. Sustain. Cities Soc. 55, 102025. https://doi.org/10.1016/j.scs.2020.102025

Jänicke, B., Holtmann, A., Kim, K.R., Kang, M., Fehrenbach, U., Scherer, D., 2018. Quantification and evaluation of intra-urban heat-stress variability in Seoul, Korea. Int. J. Biometeorol. https://doi.org/10.1007/s00484-018-1631-2

Jianan, X., Zhiyun, O., Hua, Z., Xiaoke, W., Hong, M., 2007. Allergenic pollen plants and their influential factors in urban areas. Acta Ecol. Sin. 27, 3820–3827. https://doi.org/10.1016/S1872-2032(07)60082-1

Jochner, S.C., Beck, I., Behrendt, H., Traidl-Hoffmann, C., Menzel, A., 2011. Effects of extreme spring temperatures on urban phenology and pollen production: A case study in Munich and Ingolstadt. Clim. Res. 49, 101–112. https://doi.org/10.3354/cr01022

Karl, T., Gohm, A., Rotach, M.W., Ward, H.C., Graus, M., Cede, A., Wohlfahrt, G., Hammerle, A., Haid, M., Tiefengraber, M., Lamprecht, C., Vergeiner, J., Kreuter, A., Wagner, J., Staudinger, M., 2020. Studying urban climate and air quality in the alps. Bull. Am. Meteorol. Soc. 101, E488–E507. https://doi.org/10.1175/BAMS-D-19-0270.1

Klysik, K., Fortuniak, K., 1999. Temporal and spatial characteristics of the urban heat island of Lodz, Poland. Atmos. Environ. 33, 3885–3895. https://doi.org/doi:10.1016/S1352-2310(99)00131-4

Kolokotroni, M., Ren, X., Davies, M., Mavrogianni, A., 2012. London's urban heat island: Impact on current and future energy consumption in office buildings. Energy Build. 47, 302–311. https://doi.org/10.1016/j.enbuild.2011.12.019

Kotharkar, R., Bagade, A., 2018. Evaluating urban heat island in the critical local climate zones of an Indian city. Landsc. Urban Plan. 169, 92–104. https://doi.org/10.1016/j.landurbplan.2017.08.009

Kotharkar, R., Bagade, A., Ramesh, A., 2019. Assessing urban drivers of canopy layer urban heat island: A numerical modeling approach. Landsc. Urban Plan. 190, 103586. https://doi.org/10.1016/j.landurbplan.2019.05.017

Kottek, M., Grieser, J., Beck, C., Rudolf, B., Rubel, F., 2006. World map of the Köppen-Geiger climate classification updated. Meteorol. Zeitschrift 15, 259–263. https://doi.org/10.1127/0941-2948/2006/0130

Kotthaus, S., Grimmond, C.S.B., 2014. Energy exchange in a dense urban environment - Part I: Temporal variability of long-term observations in central London. Urban Clim. 10, 261–280. https://doi.org/10.1016/j.uclim.2013.10.002

Kourtidis, K., Georgoulias, A.K., Rapsomanikis, S., Amiridis, V., Keramitsoglou, I., Hooyberghs, H., Maiheu, B., Melas, D., 2015. A study of the hourly variability of the urban heat island effect in the Greater Athens Area during summer. Sci. Total Environ. 517, 162–177. https://doi.org/10.1016/j.scitotenv.2015.02.062

Kratzer, A., 1937. Das Stadtklima. Friedr. Vieweg and Sohn Braunschweig, Braunschweig.

Kurppa, M., Nordbo, A., Haapanala, S., Järvi, L., 2015. Effect of seasonal variability and land use on particle number and $CO_2$ exchange in Helsinki, Finland. Urban Clim. 13, 94–109. https://doi.org/10.1016/j.uclim.2015.07.006

Kwok, Y.T., Schoetter, R., Lau, K.K.L., Hidalgo, J., Ren, C., Pigeon, G., Masson, V., 2019. How well does the local climate zone scheme discern the thermal environment of Toulouse (France)? An analysis using numerical simulation data. Int. J. Climatol. 39, 5292–5315. https://doi.org/10.1002/joc.6140

Lauzet, N., Rodler, A., Musy, M., Azam, M.H., Guernouti, S., Mauree, D., Colinart, T., 2019. How

building energy models take the local climate into account in an urban context – A review. Renew. Sustain. Energy Rev. 116, 109390. https://doi.org/10.1016/j.rser.2019.109390

Leconte, F., Bouyer, J., Claverie, R., 2020. Nocturnal cooling in Local Climate Zone: Statistical approach using mobile measurements. Urban Clim. 33, 100629. https://doi.org/10.1016/j.uclim.2020.100629

Leconte, F., Bouyer, J., Claverie, R., Pétrissans, M., 2017. Analysis of nocturnal air temperature in districts using mobile measurements and a cooling indicator. Theor. Appl. Climatol. 130, 365–376. https://doi.org/10.1007/s00704-016-1886-7

Leconte, F., Bouyer, J., Claverie, R., Pétrissans, M., 2015. Using Local Climate Zone scheme for UHI assessment: Evaluation of the method using mobile measurements. Build. Environ. 83, 39–49. https://doi.org/10.1016/j.buildenv.2014.05.005

Lehnert, M., Geletič, J., Husák, J., Vysoudil, M., 2015. Urban field classification by "local climate zones" in a medium-sized Central European city: the case of Olomouc (Czech Republic). Theor. Appl. Climatol. 122, 531–541. https://doi.org/10.1007/s00704-014-1309-6

Lelovics, E., Unger, J., Gal, T., Gal, C. V, Gál, T., Gál, C. V., 2014. Design of an urban monitoring network based on Local Climate Zone mapping and temperature pattern modelling. Clim. Res. 60, 51–62. https://doi.org/10.3354/cr01220

Li, X., Ratti, C., Seiferling, I., 2017. Mapping Urban Landscapes Along Streets Using Google Street View. Adv. Cartogr. GIScience 341–356. https://doi.org/10.1007/978-3-319-57336-6_24

López-Bueno, J., Díaz, J., Linares, C., 2019. Differences in the impact of heat waves according to urban and peri-urban factors in Madrid. Int. J. Biometeorol. https://doi.org/10.1007/s00484-019-01670-9

López-Bueno, J.A., Díaz, J., Sánchez-Guevara, C., Sánchez-Martínez, G., Franco, M., Gullón, P., Núñez Peiró, M., Valero, I., Linares, C., 2020. The impact of heat waves on daily mortality in districts in Madrid: The effect of sociodemographic factors. Environ. Res. 190, 109993. https://doi.org/10.1016/j.envres.2020.109993

Meier, F., Fenner, D., Grassmann, T., Otto, M., Scherer, D., 2017. Crowdsourcing air temperature from citizen weather stations for urban climate research. Urban Clim. 19, 170–191. https://doi.org/10.1016/j.uclim.2017.01.006

Menzer, O., McFadden, J.P., 2017. Statistical partitioning of a three-year time series of direct urban net $CO_2$ flux measurements into biogenic and anthropogenic components. Atmos. Environ. 170, 319–333. https://doi.org/10.1016/j.atmosenv.2017.09.049

Miao, C., Yu, S., Hu, Y., Zhang, H., He, X., Chen, W., 2020. Review of methods used to estimate the sky view factor in urban street canyons. Build. Environ. 168, 106497. https://doi.org/10.1016/j.buildenv.2019.106497

Mills, G., 2014. Urban climatology: History, status and prospects. Urban Clim. 10, 479–489. https://doi.org/10.1016/j.uclim.2014.06.004

Ministerio de Hacienda, 2019. Catastro inmobiliario. Dir. Gen. del Catastro.

Mirzaei, P.A., 2015. Recent challenges in modeling of urban heat island. Sustain. Cities Soc. 1–7. https://doi.org/10.1016/j.scs.2015.04.001

Muller, C.L., Chapman, L., Grimmond, C.S.B., Young, D.T., Cai, X.M., 2013. Toward a standardized metadata protocol for urban meteorological networks. Bull. Am. Meteorol. Soc. 94, 1161–1185. https://doi.org/10.1175/BAMS-D-12-00096.1

Núñez Peiró, M., Sánchez-Guevara, C., Neila González, F.J., 2017. Update of the Urban Heat Island of Madrid and Its Influence on the Building's Energy Simulation, in: Sustainable Development and Renovation in Architecture, Urbanism and Engineering. pp. 339–350. https://doi.org/10.1007/978-

3-319-51442-0_28

Núñez Peiró, M., Sánchez-Guevara Sánchez, C., Neila González, F.J., 2019. Source area definition for local climate zones studies. A systematic review. Build. Environ. 148, 258–285. https://doi.org/10.1016/j.buildenv.2018.10.050

Núñez Peiró, M., Sánchez-Guevara Sánchez, C., Neila González, F.J., 2018. Abrigo meteorológico para sensores ambientales. ES-2642617-B2.

Oke, T.R., 1998. An algorithm scheme to estimate hourly heat island magnitude, in: The Second Symposium on Urban Environment. American Meteorological Society, Albuquerque.

Oke, T.R., 1982. The energetic basis of the urban heat island. Q. J. R. Meteorol. Soc. 108, 1–24. https://doi.org/10.1002/qj.49710845502

Oke, T.R., Mills, G., Christen, A., Voogt, J.A., 2017a. 2 Concepts, in: Urban Climates. Cambridge University Press, pp. 14–43. https://doi.org/10.1017/9781139016476.003

Oke, T.R., Mills, G., Christen, A., Voogt, J.A., 2017b. 4 Airflow, in: Urban Climates. Cambridge University Press, pp. 77–121. https://doi.org/10.1017/9781139016476.005

Pawlak, W., Fortuniak, K., 2016. Eddy covariance measurements of the net turbulent methane flux in the city centre-results of 2-year campaign in Lodz, Poland. Atmos. Chem. Phys. 16, 8281–8294. https://doi.org/10.5194/acp-16-8281-2016

Perera, N.G.., Emmanuel, R., 2016. A "Local Climate Zone" based approach to urban planning in Colombo, Sri Lanka. Urban Clim. https://doi.org/10.1016/j.uclim.2016.11.006

Puliafito, S.E., Bochaca, F.R., Allende, D.G., Fernandez, R., 2013. Green Areas and Microscale Thermal Comfort in Arid Environments: A Case Study in Mendoza, Argentina. Atmos. Clim. Sci. 03, 372–384. https://doi.org/10.4236/acs.2013.33039

Pyrgou, A., Castaldo, V.L., Pisello, A.L., Cotana, F., Santamouris, M., 2017. Differentiating responses of weather files and local climate change to explain variations in building thermal-energy performance simulations. Sol. Energy 153, 224–237. https://doi.org/https://doi.org/10.1016/j.solener.2017.05.040

Reckien, D., Salvia, M., Heidrich, O., Church, J.M., Pietrapertosa, F., De Gregorio-Hurtado, S., D'Alonzo, V., Foley, A., Simoes, S.G., Krkoška Lorencová, E., Orru, H., Orru, K., Wejs, A., Flacke, J., Olazabal, M., Geneletti, D., Feliu, E., Vasilie, S., Nador, C., Krook-Riekkola, A., Matosović, M., Fokaides, P.A., Ioannou, B.I., Flamos, A., Spyridaki, N.A., Balzan, M. V., Fülöp, O., Paspaldzhiev, I., Grafakos, S., Dawson, R., 2018. How are cities planning to respond to climate change? Assessment of local climate plans from 885 cities in the EU-28. J. Clean. Prod. 191, 207–219. https://doi.org/10.1016/j.jclepro.2018.03.220

Renou, P.M.E., 1858. Instructions Météorologiques et Tables Usuelles, Société Météorologique de France. Paris.

Richard, Y., Emery, J., Dudek, J., Pergaud, J., Chateau-Smith, C., Zito, S., Rega, M., Vairet, T., Castel, T., Thévenin, T., Pohl, B., 2018. How relevant are local climate zones and urban climate zones for urban climate research? Dijon (France) as a case study. Urban Clim. 26, 258–274. https://doi.org/10.1016/j.uclim.2018.10.002

Roth, M., Jansson, C., Velasco, E., 2016. Multi-year energy balance and carbon dioxide fluxes over a residential neighbourhood in a tropical city. Int. J. Climatol. 37, 2679–2698. https://doi.org/10.1002/joc.4873

Rubel, F., Brugger, K., Haslinger, K., Auer, I., 2017. The climate of the European Alps: Shift of very high resolution Köppen-Geiger climate zones 1800-2100. Meteorol. Zeitschrift 26, 115–125. https://doi.org/10.1127/metz/2016/0816

Runnalls, K.E., Oke, T.R., 2006. A techique to detect microclimatic inhomogeneities in historical records of screen-level air temperature. J. Clim. 19, 959–978. https://doi.org/10.1175/JCLI3663.1

Rusel, F.A.R., 1888. Smoke in relation to fogs in London. Nature 39, 34–36. https://doi.org/10.1038/039034a0

Sánchez-Guevara, C., Núñez Peiró, M., Taylor, J., Mavrogianni, A., Neila González, J., 2019. Assessing population vulnerability towards summer energy poverty: Case studies of Madrid and London. Energy Build. 190, 132–143. https://doi.org/10.1016/j.enbuild.2019.02.024

Schatz, J., Kucharik, C.J., 2014. Seasonality of the Urban Heat Island Effect in Madison, Wisconsin. J. Appl. Meteorol. Climatol. 53, 2371–2386. https://doi.org/10.1175/JAMC-D-14-0107.1

Šećerov, I.B., Savić, S.M., Milošević, D.D., Arsenović, D.M., Dolinaj, D.M., Popov, S.B., 2019. Progressing urban climate research using a high-density monitoring network system. Environ. Monit. Assess. 191. https://doi.org/10.1007/s10661-019-7210-0

Shi, Y., Ren, C., Lau, K.K.L., Ng, E., 2019. Investigating the influence of urban land use and landscape pattern on PM2.5 spatial variation using mobile monitoring and WUDAPT. Landsc. Urban Plan. 189, 15–26. https://doi.org/10.1016/j.landurbplan.2019.04.004

Siu, L.W., Hart, M.A., 2013. Quantifying urban heat island intensity in Hong Kong SAR, China. Environ. Monit. Assess. 185, 4383–4398. https://doi.org/10.1007/s10661-012-2876-6

Skarbit, N., Stewart, I.D., Unger, J., Gál, T., 2017. Employing an urban meteorological network to monitor air temperature conditions in the "local climate zones" of Szeged, Hungary. Int. J. Climatol. https://doi.org/10.1002/joc.5023

Stewart, I.D., 2019. Why should urban heat island researchers study history? Urban Clim. 30, 100484. https://doi.org/10.1016/j.uclim.2019.100484

Stewart, I.D., Oke, T.R., 2012. Local climate zones for urban temperature studies. Bull. Am. Meteorol. Soc. 93, 1879–1900. https://doi.org/10.1175/BAMS-D-11-00019.1

Stewart, I.D., Oke, T.R., Krayenhoff, E.S., 2014. Evaluation of the "local climate zone" scheme using temperature observations and model simulations. Int. J. Climatol. 34, 1062–1080. https://doi.org/10.1002/joc.3746

Suomi, J., 2018. Extreme temperature differences in the city of Lahti, southern Finland : Intensity, seasonality and environmental drivers. Weather Clim. Extrem. 19, 20–28. https://doi.org/10.1016/j.wace.2017.12.001

Thapa Chhetri, D.B., Fujimori, Y., Moriwaki, R., 2017. Local Climate Classification and Urban Heat/Dry Island in Matsuyama Plain. J. Japan Soc. Civ. Eng. Ser. B1 (Hydraulic Eng. 73, I_487-I_492. https://doi.org/10.2208/jscejhe.73.i_487

Thomas, G., Sherin, A.P., Ansar, S., Zachariah, E.J., 2014. Analysis of Urban Heat Island in Kochi, India, Using a Modified Local Climate Zone Classification. Procedia Environ. Sci. 21, 3–13. https://doi.org/10.1016/j.proenv.2014.09.002

Toparlar, Y., Blocken, B., Maiheu, B., van Heijst, G.J.F., 2017. A review on the CFD analysis of urban microclimate. Renew. Sustain. Energy Rev. 80, 1613–1640. https://doi.org/10.1016/j.rser.2017.05.248

Unger, J., Gál, T., Csépe, Z., Lelovics, E., Gulyás, Á., 2015. Development, data processing and preliminary results of an urban human comfort monitoring and information system. Idojaras 119, 337–354.

Universidad Politécnica de Madrid, 2014. MODIFICA Project: Predictive model for dwellings energy performance under the urban heat island effect [WWW Document]. Minist. Econ. Compet. URL BIA2013-41732-R

Valach, A.C., Langford, B., Nemitz, E., Mackenzie, A.R., Hewitt, C.N., 2015. Seasonal and diurnal trends in concentrations and fluxes of volatile organic compounds in central London. Atmos. Chem. Phys. 15, 7777–7796. https://doi.org/10.5194/acp-15-7777-2015

Velasco, E., 2018. Go to field, look around, measure and then run models. Urban Clim. 24, 231–236. https://doi.org/10.1016/j.uclim.2018.04.001

Velasco, E., Perrusquia, R., Jiménez, E., Hernández, F., Camacho, P., Rodríguez, S., Retama, A., Molina, L.T., 2014. Sources and sinks of carbon dioxide in a neighborhood of Mexico City. Atmos. Environ. 97, 226–238. https://doi.org/10.1016/j.atmosenv.2014.08.018

Verdonck, M.L., Demuzere, M., Hooyberghs, H., Beck, C., Cyrys, J., Schneider, A., Dewulf, R., Van Coillie, F., 2018. The potential of local climate zones maps as a heat stress assessment tool, supported by simulated air temperature data. Landsc. Urban Plan. 178, 183–197. https://doi.org/10.1016/j.landurbplan.2018.06.004

Willers, S.M., Jonker, M.F., Klok, L., Keuken, M.P., Odink, J., van den Elshout, S., Sabel, C.E., Mackenbach, J.P., Burdorf, A., 2016. High resolution exposure modelling of heat and air pollution and the impact on mortality. Environ. Int. 89–90, 102–109. https://doi.org/10.1016/j.envint.2016.01.013

WMO, 2017a. Guide to Meteorological Instruments and Methods of Observation (WMO No. 8), World Meteorological Organization. Geneva, Switzerland.

WMO, 2017b. Guide to the Global Observing System (WMO No. 488). Geneva, Switzerland.

Yagüe, C., Zurita, E., Martinez, A., 1991. Statistical analysis of the Madrid urban heat island. Atmos. Environ. 25, 327–332. https://doi.org/10.1016/0957-1272(91)90004-X

Yang, P., Ren, G., Liu, W., 2013. Spatial and Temporal Characteristics of Beijing Urban Heat Island Intensity. J. Appl. Meteorol. Climatol. 52, 1803–1816. https://doi.org/10.1175/JAMC-D-12-0125.1

Yang, X., Chen, Y., Peng, L.L.H., Wang, Q., 2020a. Quantitative methods for identifying meteorological conditions conducive to the development of urban heat islands. Build. Environ. 178, 106953. https://doi.org/10.1016/j.buildenv.2020.106953

Yang, X., Peng, L.L.H., Chen, Y., Yao, L., Wang, Q., 2020b. Air humidity characteristics of local climate zones: A three-year observational study in Nanjing. Build. Environ. 171, 106661. https://doi.org/10.1016/j.buildenv.2020.106661

Yang, X., Yao, L., Jin, T., Peng, L.L.H., Jiang, Z., Hu, Z., Ye, Y., 2018. Assessing the thermal behavior of different local climate zones in the Nanjing metropolis, China. Build. Environ. 137, 171–184. https://doi.org/10.1016/j.buildenv.2018.04.009

Yuan, C., Adelia, A.S., Mei, S., He, W., Li, X.X., Norford, L., 2020. Mitigating intensity of urban heat island by better understanding on urban morphology and anthropogenic heat dispersion. Build. Environ. 176, 106876. https://doi.org/10.1016/j.buildenv.2020.106876

Zahumensky, I., 2004. Guidelines on Quality Control Procedures for Data from Automatic Weather Stations. Geneva, Switzerland.

Zhou, X., Okaze, T., Ren, C., Cai, M., Ishida, Y., Watanabe, H., Mochida, A., 2020. Evaluation of urban heat islands using local climate zones and the influence of sea-land breeze. Sustain. Cities Soc. 55, 102060. https://doi.org/10.1016/j.scs.2020.102060

Zuvela-Aloise, M., 2017. Enhancement of urban heat load through social inequalities on an example of a fictional city King's Landing. Int. J. Biometeorol. 61, 527–539. https://doi.org/10.1007/s00484-016-1230-z